\documentclass[fleqn,usenatbib,twocolumn]{mnras}

\usepackage[T1]{fontenc}

\DeclareRobustCommand{\VAN}[3]{#2}
\let\VANthebibliography\thebibliography
\def\thebibliography{\DeclareRobustCommand{\VAN}[3]{##3}\VANthebibliography}

\usepackage{graphicx}	
\usepackage{float}
\usepackage{amsmath}	
\usepackage{amssymb}	
\usepackage{mathrsfs}
\usepackage{xcolor}

\newcommand{\DzerocovaryBeta}{(26.06\beta-5.54)\pm0.005}
\newcommand{\betaLocalRel}{0.4036\pm0.0016}

\newcommand{\completenessMatchMPAJHU}{0.938}
\newcommand{\completenessVp}{0.977}
\newcommand{\completenessVpSD}{0.029}

\newcommand{\kms}{km\;s$^{-1}$}
\newcommand{\unitP}{$h^{-3}$\,Mpc$^{3}$\,km$^{2}$\,s$^{-2}$}
\newcommand{\unitNBAR}{$h^{3}$ Mpc$^{-3}$}
\newcommand{\fsigma}{0.459$^{+0.068}_{-0.069}$}
\newcommand{\fsigmaLZ}{0.416$^{+0.074}_{-0.076}$}
\newcommand{\fsigmaHZ}{0.526$^{+0.133}_{-0.148}$}
\newcommand{\IncreaseErrorRSD}{2.1}

\newcommand{\zeff}{0.0959}

\newcommand{\totnumPVCatalog}{229,890}
\newcommand{\IncreaseVolumeSDSSFP}{2.4}
\newcommand{\IncreaseVolumesixdFGSv}{6.7}
\newcommand{\IncreaseFPnum}{2.5}  

\title[Momentum Power Spectrum to $z$=0.2]{Momentum power spectrum of
  SDSS galaxies  by massE cosmic ruler: \IncreaseErrorRSD$\times$ improvement in measure of growth rate}

\author[Y. Shi et al.]{
Yong Shi,$^{1,2}$\thanks{E-mail: yong@nju.edu.cn}, Pengjie Zhang$^{3,4}$, Shude Mao$^{5}$, Qiusheng Gu$^{1,2}$
\\
$^{1}$School of Astronomy and Space Science, Nanjing University, Nanjing 210093, China.\\
$^{2}$Key Laboratory of Modern Astronomy and Astrophysics (Nanjing University), Ministry of Education, Nanjing 210093, China.\\
$^{3}$Department of Astronomy, School of Physics and Astronomy, Shanghai Jiao Tong University,
Shanghai, 200240, China.\\
$^{4}$Tsung-Dao Lee Institute, Shanghai Jiao Tong University, Shanghai 200240, China.\\
$^{5}$Department of Astronomy, Tsinghua University, Beijing 100084, China.\\
}

\date{Accepted XXX. Received YYY; in original form ZZZ}

\pubyear{2023}

\begin{document}
\label{firstpage}
\pagerange{\pageref{firstpage}--\pageref{lastpage}}
\maketitle

\begin{abstract}

\noindent Peculiar motion  of galaxies probes the  structure growth in
the Universe.  In this study we employ the galaxy stellar mass-binding
energy (massE) relation with only two nuisance parameters to build the
largest  peculiar-velocity   (PV)  catalog  to  date,   consisting  of
\totnumPVCatalog\; ellipticals  from the  main galaxy sample  (MGS) of
the Sloan Digital Sky Survey  (SDSS).  We quantify the distribution of
the  massE-based   distances  in   individual  narrow   redshift  bins
($dz$=0.005), and  then estimate the  PV of  each galaxy based  on its
offset from  the Gaussian  mean of  the distribution.  As demonstrated
with  the Uchuu-SDSS  mock data,  the  derived PV  and momentum  power
spectra are insensitive to accurate  calibration of the massE relation
itself, enabling  measurements out to  a redshift of 0.2,  well beyond
the current limit of $z$=0.1 using other galaxy scaling laws.  We then
measure the  momentum power spectrum  and demonstrate that  it remains
almost unchanged if varying significantly the redshift bin size within
which the distance is measured, as  well as the intercept and slope of
the massE  relation, respectively.  By  fitting the spectra  using the
perturbation theory model with  four free parameters, $f\sigma_{8}$ is
constrained  to   $f\sigma_{8}$=\fsigma\;  over  ${\Delta}z$=0.02-0.2,
\fsigmaLZ\,   over    ${\Delta}z$=0.02-0.1   and    \fsigmaHZ\,   over
${\Delta}z$=0.1-0.2.      The     error    of     $f\sigma_{8}$     is
\IncreaseErrorRSD\,  times smaller  than  that by  the redshift  space
distortion  (RSD)  of  the   same  sample.  A  Fisher-matrix  forecast
illustrates that the constraint  on $f\sigma_{8}$ from the massE-based
PV can potentially exceed that from  the stage-IV RSD in late universe
($z$$<$0.5).

\end{abstract}

\begin{keywords}
galaxies: general - galaxies: kinematics and dynamics - (cosmology:) large-scale structure of Universe- cosmology: observations - (cosmology:) distance scale
\end{keywords}



\section{Introduction} \label{sec_intro}

The  large-scale  structure  of  galaxies, as  a  consequence  of  the
structure growth  of matter in  the expanding Universe, is  a powerful
way to  understand the  cosmos \citep{Peebles80}.  The  average excess
number of galaxies  surrounding a galaxy as a  function of separation,
i.e., the  correlation function  of the galaxy  number density  or its
Fourier analog--power  spectrum, has been widely  used to characterize
the  large scale  structure \citep{Kaiser87,  Eisenstein05, Beutler12,
  Alam17,  Alam21}.  This  includes  studying  the  baryonic  acoustic
oscillation around a separation of 100 Mpc/$h$, as well as examining the
redshift  space   distortion  and  the  full   broad-band  correlation
function.

In addition to the number density, galaxies experience peculiar motion
under the gravitational potential of  matter. Unlike the galaxy number
density that is a biased tracer of matter, peculiar velocity (PV) is a more
direct tracer at least on linear scales \citep{Koda14, Howlett19}.
When the density perturbation is   small,   the   over-density   evolves  with   cosmic   time   as:
$\delta$(\mathbfit{r},$a$)=$\rho$(\mathbfit{r},$a$)/
$\bar{\rho}$(\mathbfit{r},$a$)-1 $\propto$ $D(a)$      where
\mathbfit{r} is  a comoving  coordinate, $a$ is  the scale  factor and
$D(a)$  is the  linear growth  factor.  The  continuity equation  then
gives
\begin{equation}
  \mathbf{\nabla} {\cdot} \mathbfit{v}(\mathbfit{r},a)  = - a\frac{ {\rm d}\delta(\mathbfit{r},a)}{{\rm d}t}
  =  -a^{2}H(a)f(a)\delta(\mathbfit{r},a),
\end{equation}
where the  Hubble constant $H(a)={\rm  d}a/{\rm d}t/a$ and  the growth
rate $f(a)$= ${\rm dln}D(a)/{\rm  dln}a$.  The equation indicates that
on linear scales the PV divergence is directly related
to the  growth rate  $f(a)$ of matter. Under the  General Relativity
(GR), $f(a)=\Omega_{\rm m}(a)^{0.55}$ \citep{Lahav91, Linder07}, and  any deviation from it could
imply the break  down of GR. By taking  the fact that
the Fourier transform  of a quantity is proportional  to the transform
of its divergence, $\mathscr{F}$($\mathbf{\nabla}{\cdot}$\mathbfit{v}) =$i\mathbfit{k}{\cdot}$ $\mathscr{F}$(\mathbfit{v}).
As a result, in addition to the density field, the  PV field of a large
scale structure offers a complementary  way to constrain the structure
growth  or test  GR by  quantifying the  velocity power  spectrum.  In
practice,  we only  measure the  PV  along the  line of
sight  to  each observed  galaxy,  which  thus gives  a  line-of-sight
mass-weighted velocity power spectrum or momentum power spectrum.

With a PV as denoted by $v_{\rm p}$, the observed redshift $z_{\rm obs}$ follows
\begin{equation}\label{eqn_zobs_zcos}
  1+z_{\rm obs} = (1+z_{\rm cos})(1+\frac{v_{\rm p}}{c}),
\end{equation}
where $z_{\rm cos}$ is the cosmic redshift and $c$ is the speed of light. Measurements of PV require
measurements of observed redshift as well as cosmological-independent measurements of
distances that infer $z_{\rm cos}$. The latter has been achieved through galaxy scaling law
including the Tully-Fisher relation for spiral galaxies \citep{Tully77},  the fundamental plane (FP) \citep{Dressler87} and surface brightness
  fluctuation for elliptical galaxies \citep{Tonry88, Tonry01} as well as SN Ia \citep{Riess98}.

Studies of momentum power spectrum has significant development both in
theories   and  observations.    Through  distribution   function  and
perturbation theory, non-linear effects on the momentum power spectrum
has been characterized  analytically \citep{Vlah12, Vlah13, Okumura14,
  Saito14}, which makes possible the comparison to observations over a
full band  of spatial  scales. In observations,  the technique  of the
density   power   spectrum   \citep{Feldman94}  has   been   developed
\citep{Koda14,  Hand17,  Hand18,  Howlett19} and  applied  to  several
PV catalog  to  estimate the  momentum power  spectrum
\citep[e.g.][]{Qin19}.   These  measurements,   along  with   velocity
correlation   function   or    velocity-density   cross   correlations
\citep{Johnson14,   Adams17,   Adams20,  Wang21, Turner23, Lai23},   have   produced
reasonable constraints on the growth rate.

Exiting  PV  surveys based  on galaxy  scaling law  are
limited to  very low  redshift, i.e.,  not beyond  a redshift  of 0.1,
while those with  SN Ia can go  to higher redshift but  its low number
statistics  make it  not  competitive.  For  example,  the Two  Micron
All-Sky Survey (2MASS) Tully-Fisher survey  covers a redshift range up
to   0.033   with   about   2000   objects   \citep{Masters08}.    The
Six-degree-Field  Galaxy   Survey  peculiar-velocity  (6dFGSv)   is  a
FP-based PV survey up to a redshift of 0.053 with about
9000  objects \citep{Magoulas12, Springob14}.   The  SDSS FP-based  peculiar
velocity   sample  extends   to  a   redshift  of   0.1  with   $\sim$
3.4$\times$10$^{4}$    objects   \citep{Howlett22}.    {\it Cosmicflows-IV}
data-set compiles a total number of $\sim$ 5.5$\times$10$^{4}$ objects
with all  kinds of distance  calibration \citep{Tully22}, including
those with surface brightness fluctuation and SN Ia etc. 

Although  the measurement  error of  PV  increases  with  redshift,
both  the increasing  volume  at  a  larger
redshift and  the sampling  on larger spatial  scales can  improve the
constraint  on  the  growth   rate  $f$  \citep[e.g.][]{Koda14}.   The
Tully-Fisher  relation  requires   spatially-resolved  measurement  of
kinematic maps  which is difficult  beyond $z$=0.1.  Although  FP only
requires single-fiber measurement of  velocity dispersion, the FP relation contains
  three free parameters plus an additional 
nuisance    parameter that quantifies the offset from the FP plane for galaxies with different properties
  \citep{Magoulas12, Howlett22}.

In  \citet{Shi21}, the  stellar mass  of a  galaxy has  been found  to
tightly correlate with the galaxy  binding energy within the effective
radius, which is referred as the massE relation.  This relation offers
a new cosmic ruler with only two nuisance parameters \citep{Shi22}, in
contrast to the FP that has four nuisance parameters.  The application
of  the massE  cosmic ruler  to ellipticals  of the  SDSS main  galaxy
sample (MGS)  offer distance  measurement with an  accuracy as  low as
0.35\% for the redshift range  of 0.05-0.2, which proves the existence
of dark  energy at  7-$\sigma$ under  flat-$\Lambda$CDM \citep{Shi22}.
In this study, we use the  massE to measure PV of the
SDSS MGS elliptical sample and  obtain the momentum power spectrum out
to $z$=0.2.   Throughout the study, we adopt the Planck 2018 flat
$\Lambda$-CDM cosmology \citep{Planck20} with $h$=0.6736, $\Omega_{\rm
  m}$=0.3153 and $\Omega_{\rm b}$ =0.0493 and $\sigma_{8}$=0.811.

\section{Measurements of peculiar velocities of the SDSS MGS-elliptical sample}

\subsection{The massE cosmic ruler as the distance estimator}

The distance  measurement of the  SDSS MGS elliptical sample  with the
massE cosmic  ruler is detailed  in \citet{Shi22}. Briefly,  the massE
ruler estimates the angular diameter  distance of an elliptical galaxy with stellar mass well
above 10$^{8}$ $M_{\odot}$
as:
\begin{eqnarray}\label{eqn_distance_ruler}
  \frac{D_{\rm A}}{\rm Mpc} & = &  S_{D_{0}} \frac{D_{0}}{\rm Mpc}\left[(1+z_{\rm cos})^{-4\beta}(\frac{\sigma_{\rm e}}{\rm km\,s^{-1}})(\frac{M_{\rm star,1Mpc}}{M_{\odot}})^{-\beta}\right.  \nonumber   \\ 
  & &  \left.(\frac{R_{\rm e,as}}{\rm arcsec})^{0.25}\right]^{1/(2\beta-0.25)},  
\end{eqnarray}
where  $R_{\rm  e,as}$ is
the apparent  effective radius in  arcsec, 
$\sigma_{\rm  e}$ is the velocity dispersion within $R_{\rm  e,as}$, 
$M_{\rm star,1Mpc}$  is the galaxy
stellar mass placed at  a distance of 1  Mpc and $z_{\rm cos}$ is the cosmic
redshift of the galaxy. Besides the above four observables,
the two  nuisance parameters in Equation~\ref{eqn_distance_ruler} are $\beta$ and
$S_{D_{0}}$, where $\beta$ 
is the slope of the massE relationship, and $S_{D_{0}}$ is the intercept of the relation and is related to the absolute
calibration of observables and is degenerate with the local Hubble constant $H_{0}$. By
fitting to the low-redshift sample
compiled in  \citet{Shi21}, \citet{Shi22} estimated:
\begin{equation}\label{eqn_beta}
  \beta  = {\betaLocalRel}
\end{equation}
and
\begin{equation}\label{eqn_D0_beta}
  {\rm ln} ({D_{0}}/{\rm Mpc})={ \DzerocovaryBeta}.
\end{equation}
The angular diameter distance shown in the above equation can be converted
the luminosity distance through $D_{\rm L}$=$D_{\rm A}$(1+z)$^{2}$. For a flat
cosmology, the angular distance is related to 
the comoving distance through $D_{\rm c}$=$D_{\rm A}$(1+z).

In practice,  the observed  redshift $z_{\rm  obs}$ is
  used  to  approximate $z_{\rm  cos}$.  Although  the effect  on  the
  distance is negligible, the effect on the PV is noticeable as shown in
  Equation~\ref{eqn_vp_eta}. By assuming $z_{\rm  cos}$=$z_{\rm  obs}$, the shift in
  the comoving distance can be derived as:
  \begin{eqnarray}\label{eqn_kappa}
  \Delta {\rm ln}(D_{\rm c})|_{(z_{\rm cos}=z_{\rm obs})} & = & \Delta {\rm ln}(D_{\rm A})|_{(z_{\rm cos}=z_{\rm obs})} + {\rm ln}(\frac{1+z_{\rm cos}}{1+z_{\rm obs}}) \nonumber \\
  & = & (\frac{-4\beta}{2\beta-0.25}+1)\;{\rm ln}(\frac{1+z_{\rm cos}}{1+z_{\rm obs}}) \nonumber \\
  & = & (\frac{4\beta}{2\beta-0.25}-1)\;{\rm ln}(1+\frac{v_{\rm p}}{c}) \nonumber \\
  & \approx & (\frac{2\beta+0.25}{2\beta-0.25})\frac{v_{\rm p}}{c} \\
  & = & {\kappa}v_{\rm p},
\end{eqnarray}
  in which we define $\kappa$=$(\frac{2\beta+0.25}{2\beta-0.25})\frac{1}{c}$, and assume that $v_{\rm p}$ is a small number
  as compared to the speed of light.

As shown later, the PV is not sensitive to the accurate calibration
of the local massE relationship, i.e., $\beta$ and $S_{D_{0}}$ can vary
significantly while the derived PV and momentum power spectrum remains almost unchanged.

\subsection{The MGS-elliptical sample}\label{MGS_ellip}

\begin{table}
\tiny
\begin{center}
\caption{\label{tab_sel_sample} The steps to  select the sample.}
\begin{tabular}{llllllllllllll}
\hline
steps & sample selection    & num. of objects & completeness \\
\hline
1 & MGS parent sample (LSS safe-0)           &     559,391   & \\
2 & cross match with MPA-JHU                 &  524,910     &  0.938 \\
3 & MGS-elliptical (see \S~\ref{MGS_ellip})  &     238,539   & \\
4 & MGS-elliptical PV catalog                &        229,890     &  0.964  \\
5 & PV catalog in North Cap                 &  204,327   &  \\
  & (for mom. power spectrum)                &                     &     \\
\hline
\end{tabular}\\
\end{center}
\end{table}

The   procedure   to   construct   the  PV   catalog   is   shown   in
Table~\ref{tab_sel_sample} that  lists the number of  objects for each
step.   We  start with  the  SDSS  Large-Scale-Structure (LSS)  safe-0
catalog  \citep{Blanton05} whose  completeness and  selection function
have                                                              been
quantified        \footnote{http://sdss.physics.nyu.edu/vagc/lss.html,
http://sdss.physics.nyu.edu/lss/dr72/}. The redshift in the catalog is
converted  from the  heliocentric  frame to  the  CMB frame  following
\citet{Fixsen96}.

To define  the elliptical  sub-sample, we  first cross-match  with the
MPA-JHU  catalog with  a  search radius  of 1.5  arcsec  to query  the
stellar   mass,  Petrosion   half-light   radius  in   the  $r$   band
($\texttt{PETROR50{\_}R}$), velocity dispersion ($\texttt{V{\_}DISP}$)
and  the  fraction  of  de  Vaucouleurs  component  in  the  $r$  band
($\texttt{FRACDEV{\_}R}$) from the MPA-JHU catalog \citep{Kauffmann03,
  Brinchmann04}.    The   fraction   of  the   successful   match   is
\completenessMatchMPAJHU, which will be  included in the galaxy weight
when calculating the momentum power spectrum.  The SDSS MGS-elliptical
sample   is  then   defined  as   $\texttt{FRACDEV{\_}R}$  $>$   0.80,
$\texttt{PETROR50{\_}R}$ > 0, along  with $\texttt{V{\_}DISP}$ > 0 and
0.02 $<$  $z_{\rm obs}$ $<$  0.2. The  lower redshift limit is adopted  so that
even in  the lowest redshift  bin of  [0.02, 0.021], 95\%  of galaxies
still have stellar masses larger than 3$\times$10$^{8}$ M$_{\odot}$ to
ensure    the   validity    of   the    distance   calculation    with
Equation~\ref{eqn_distance_ruler}. The upper redshift  limit is set to
ensure a reasonable  high galaxy  number density.  The footprint  of the
MGS-elliptical sample is shown in Figure~\ref{footprint}(a).

\subsection{Measurements of peculiar velocities}\label{sec_measure_vpec}

\begin{figure}
  \begin{center}
        \includegraphics[scale=0.75]{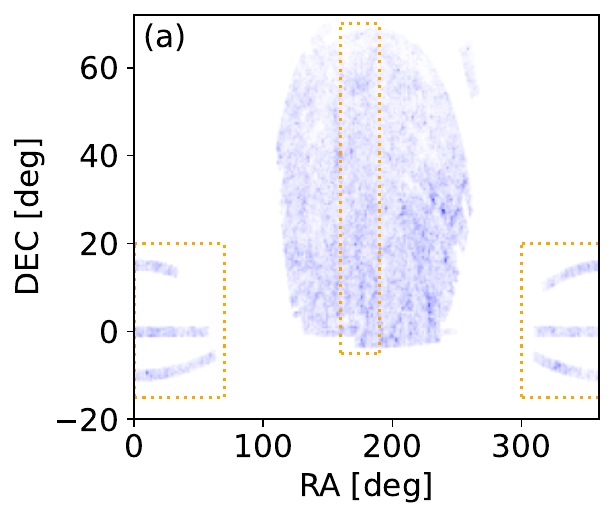}
    \includegraphics[scale=0.75]{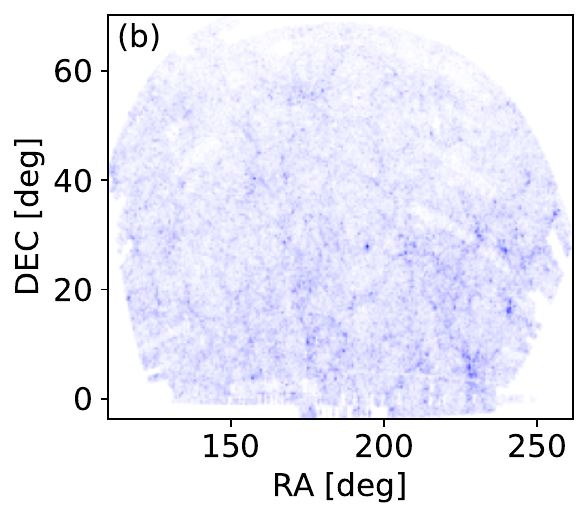}
    \caption{\label{footprint} {\bf (a),} the footprint of our massE-based SDSS MGS-PV catalog. Galaxies in three
      dotted boxes are used to measure the velocity zero point. {\bf (b),}
    the north cap portion of the PV catalog for the momentum power spectrum measurement.}
\end{center}
\end{figure}

\begin{figure*}
  \begin{center}
    \includegraphics[scale=0.5]{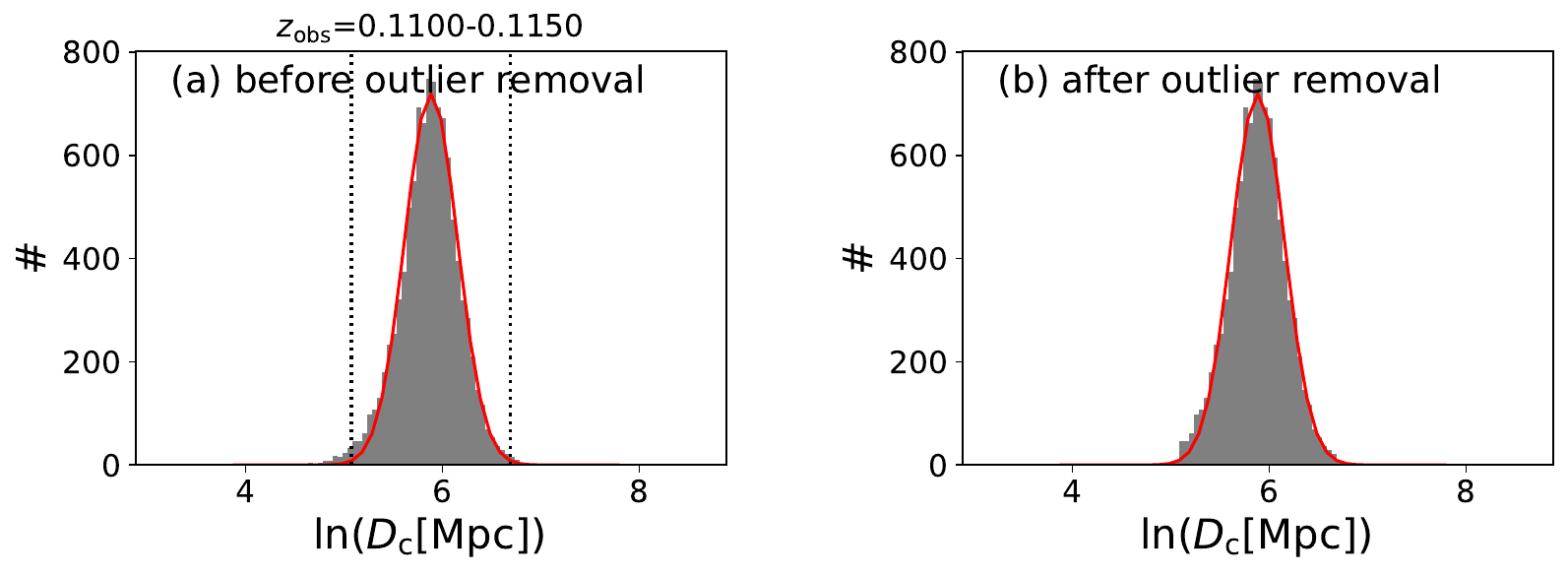}
    \caption{\label{lnobsDc_SDSS} {\bf (a)}, the distribution of the
      massE-based comoving distance
      in one redshift bin. The solid red curve is the best-fit
      Gaussian distribution and two dotted lines mark the three standard deviations. {\bf (b)}, the new distribution after removing outliers as detailed in \S~\ref{sec_measure_vpec}. }
\end{center}
\end{figure*}

\begin{figure}
  \begin{center}
    \includegraphics[scale=0.55]{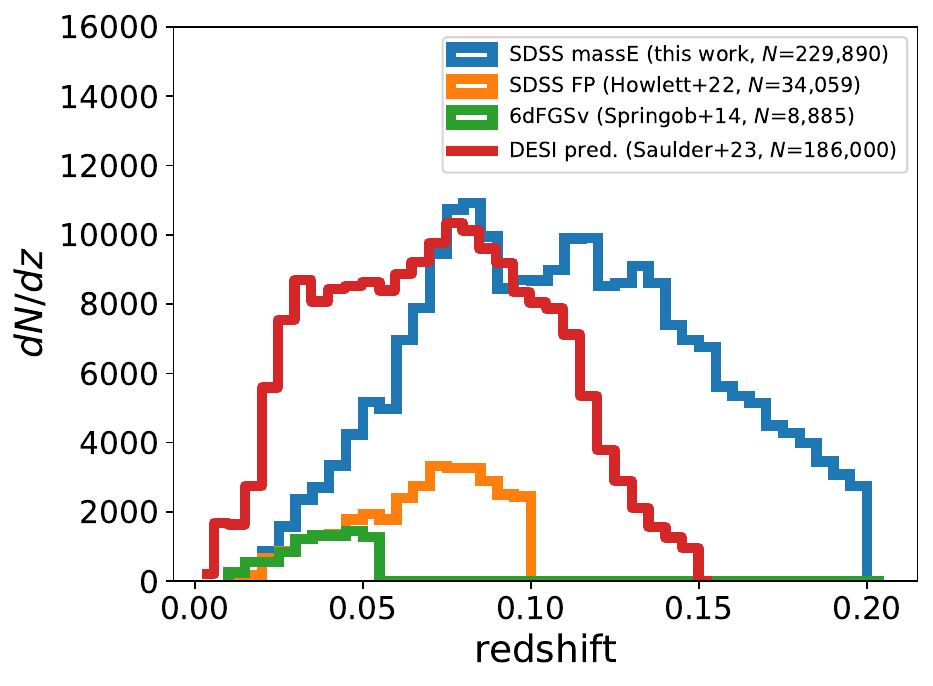}
    \caption{\label{num_PVcatalog} The number of objects as a function of redshift in
      our SDSS massE-based PV catalog, compared to those in other catalogs using
      FP or/and Tully-Fisher in the literature. }
\end{center}
\end{figure}

\begin{figure}
  \begin{center}
    \includegraphics[scale=0.7]{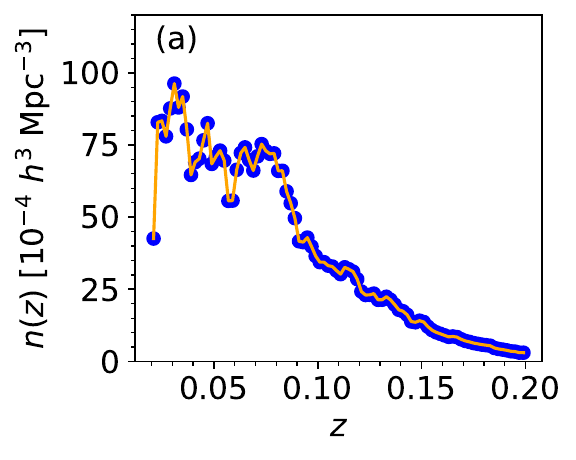}
    \includegraphics[scale=0.7]{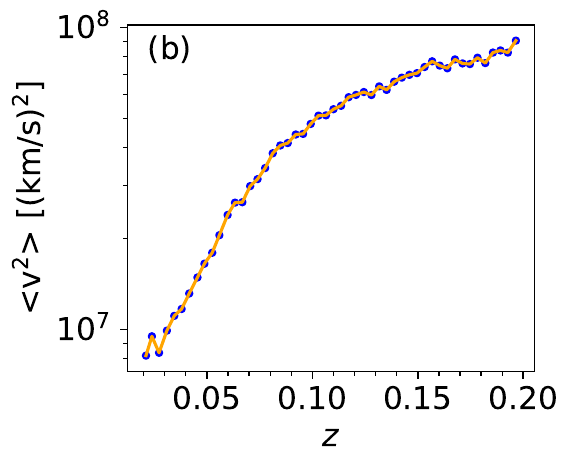}    
    \caption{\label{nofz_v2ofz} {\bf (a)}, the number density of galaxies in
      our SDSS massE-based PV catalog, where symbols are the data and curves
      are the interpolated one. {\bf (b)}, the PV variance as a function of
      redshift, where symbols are the data and the curve is the interpolated one.}
\end{center}
\end{figure}

To  measure  PV  for  the   MGS-elliptical  sample,  we  first  derive
quantities     as     needed     by    the     distance     calculator given by
Equation~\ref{eqn_distance_ruler}. As  detailed in  \citet{Shi22}, the
apparent size $R_{\rm e, as}$ is derived from the Petrosian half-light
radius $\texttt{PETROR50{\_}R}$ by adopting  the cosmological model of
the  MPA-JHU  catalog,  and  the flux-like  $M_{\rm  star,  1Mpc}$  is
converted  from the  stellar mass  in the MPA-JHU  catalog using  the same
cosmological model.  The velocity dispersion within the fiber
aperture from that catalog is corrected to the effective radius
to give $\sigma_{\rm e}$.

In practice it  is not recommended to estimate  $v_{\rm p}$ directly
from    Equation~\ref{eqn_zobs_zcos},    because   the    log-normal
distribution of the distance errors would make errors of $v_{\rm p}$
non-Gaussian, thus  complicating the error analysis  of momentum power
spectra and cosmological parameter  constraints. In previous studies
\citep{Adams17}, it  has been demonstrated  that $v_{\rm p}$  can be
estimated from  the logarithmic  ratio of  the comoving  distance at
$z$=$z_{\rm  obs}$ and  that at  $z$=$z_{\rm cos}$,  i.e., $\eta_{\rm gal}$  =
ln($\frac{D_{\rm c}(z_{\rm obs})}{D_{\rm c}(z_{\rm cos})}$):
\begin{equation}\label{eqn_vp_eta_old}
 {v_{\rm p}} =  \eta_{\rm gal}/\alpha,
\end{equation}
where
\begin{equation}
 \alpha = \frac{1+z_{\rm obs}}{D_{\rm c}(z_{\rm obs})H(z_{\rm obs})}, 
\end{equation}
with requirements of $v_{\rm p}$/c $\ll$ 1 and $\alpha{v_{\rm p}}$ $\ll$
1.  The  typical $v_{\rm  p}$ is  about 300 \kms\;  so that  the first
condition is  satisfied.  The second  condition requires  $z_{\rm
  obs}$ to be reasonably larger than 0,  e.g.  $\alpha{v_{\rm p}}$ $\sim$ 0.05 at
$z_{\rm  obs}$=0.02 and  $v_{\rm p}$=300  \kms.

In the above equation, the true  comoving distance of a galaxy,
i.e., $D_{\rm c}(z_{\rm cos})$, is  measured directly from  the massE
cosmic ruler. In practice we replace $z_{\rm cos}$ with
  $z_{\rm obs}$ in the massE equation, which causes a systematic shift in ln($D_{\rm c}(z_{\rm cos})$) as
  derived in Equation~\ref{eqn_kappa}. To account for this shift, an updated equation to derive
  PV for the massE cosmic ruler is given as 
\begin{equation}\label{eqn_vp_eta}
{v_{\rm p}} =  \eta_{\rm gal}/(\alpha+\kappa),
\end{equation}
where $\kappa$ (defined in Equation~\ref{eqn_kappa}) dominates over $\alpha$ at increasing redshift and reduces
PV errors
accordingly compared to Equation~\ref{eqn_vp_eta_old}.

The remaining key is  to obtain unbiased estimate of the
comoving  distance  at the  observed  redshift  of the  galaxy,  i.e.,
$D_{\rm  c}(z_{\rm obs})$.   Our  overall strategy  is  to divide  the
interested  redshift  range  into  small  bins  and  estimate  $D_{\rm
  c}(z_{\rm obs})$ therein. The redshift bin size needs to balance two
facts.  On the one hand, galaxies  in each  redshift bin are  composed of
those located at slightly lower true  distances with $v_{\rm p}$ >
0 and those located at slightly  larger distances with $v_{\rm p}$
< 0.  As a result, an  increasing redshift bin size  contains a larger
number of galaxies to converge their mean true distance ${\langle}{\rm
  ln}(D_{\rm  c}(z_{\rm  cos})){\rangle}$ to  the distance  at $z_{\rm
  obs}$, i.e.,  ln($D_{\rm c}(z_{\rm  obs})$). On the other  hand, the
redshift bin cannot  be too large so that $D_{\rm  c}(z_{\rm obs})$ is
no longer a constant over the bin.  

Our  fiducial measurements  adopt  a redshift  bin  size of  $d{z_{\rm
    obs}}$=5$\times$10$^{-3}$.   In each  redshift  bin,  as shown  in
Figure~\ref{lnobsDc_SDSS}(a) for an  example, the ln($D_{\rm c}(z_{\rm
  cos})$) distribution  is more or  less normal.  But because  a small
fraction  of  the  sample  has failed  measurements  in  the  velocity
dispersion or galaxy sizes, there are outliers.  We remove outliers in
the  following  ways:  for  those  within  three  times  the  standard
deviation of the best-fit Gaussian, we  include all of them; for those
outside,  we randomly  include  part of  them so  that  the number  of
objects follows the  Gaussian expectation at its  ln($D_{\rm c, z_{\rm
    obs}}$)    bin.     The    new   distribution    is    shown    in
Figure~\ref{lnobsDc_SDSS}  (b).   The   exclusion  of  these  outliers
decreases  the sample  completeness  and increases  the  error of  the
momentum power spectrum. The $\eta_{\rm bin}$ is then defined as
  the minus of the  offset of a galaxy from the  Gaussian mean in that
  redshift bin. Because the redshift bin size is not infinitely small,
  we correct additional small offset  between the observed redshift and
  the mean redshift  of galaxies within the bin:
  \begin{equation}
\eta_{\rm gal} = \eta_{\rm bin} + c\alpha(z_{\rm obs}-{\langle}z{\rangle})/(1+z_{\rm  obs}),
    \end{equation}
where ${\langle}z{\rangle}$ is  the mean  redshift of galaxies  within the
  given redshift  bin.   Besides the fiducial  redshift bin  size, we
show  that  if  varying  $dz_{\rm  obs}$  from  1$\times$10$^{-3}$  to
1$\times$10$^{-2}$, the result remains unchanged as discussed later.

Our  final  PV  catalog,   listed  in  Table~\ref{tab_sel_sample}  and
illustrated in Figure~\ref{num_PVcatalog}, contains \totnumPVCatalog\;
galaxies, representing the largest number of objects in any PV catalog
thus far.   It is seven times  larger than the previously  largest one
\citep{Howlett22} and even  exceeds the expected number  of PV objects
generated by  applying the  FP plus Tully-Fisher  methods to  the DESI
survey \citep{Saulder23}. At $z$ $<$ 0.1 we include the whole  SDSS MGS
  early-type  galaxies  with few  selection,  while  the SDSS-FP sample has
  additional  selections to  have higher  PV precision  for individual
  galaxies,  such  as a  brighter  flux  limit, galaxy inclination angle,  H$\alpha$
  equivalent width etc. As a result,  our sample at $z<$0.1 contains a
  factor  of \IncreaseFPnum\,  more galaxies  than  the SDSS-FP  one.
Another important difference is because of our  ability to
apply the massE method  up to a redshift of 0.2,  which is well beyond
the  limit  ($\sim$0.1)  of  the Tully-Fisher  and  FP  methods.   The
redshift limit of the latter two methods is not a rigid boundary based
on the first principle, but  rather derived from empirical usage.  The
Tully-Fisher    method     requires    spatially-resolved    kinematic
measurements, limiting its  redshift range.  Although the  FP can rely
on single fiber  measurements, its zero point is a  function of galaxy
properties  \citep{Magoulas12,   Howlett22}.  While
adjustments  can  be  made  to  the FP  zero  point  based  on  galaxy
properties,  even a  small remaining  systematic shift  may lead  to
a large offset in the PV estimation at relatively high redshifts.
On  the hand,  the  massE  contains solely  two  free parameters  that
correspond  to the  intercept  and slope  of  the massE  relationship,
respectively. Consequently,  we divide the redshift  range into narrow
individual  bins,  whereby PV  is  measured  by  the offset  from  the
Gaussian  mean  of  galaxies in narrow redshift bins,
largely removing the  necessity  of  accurately
measuring the zero  point throughout the entire redshift  range.
We will present  the investigation using mock
  data-sets to support the above statement in \S~\ref{sec_comp_vpec_mock}. 
We do
not see  any apparent limit for  massE to extend beyond  a redshift of
0.2. However, given the rapid drop in the galaxy number density of the
SDSS MGS toward  higher redshift, in this study, we  define our sample
to be below a redshift of 0.2.

For the momentum power spectrum measurement, we select those that have
completeness \texttt{fgotten} $\geq$ 0.9 and lie within the contiguous
region of the North Galactic  cap. The corresponding footprint area is
6725 deg$^{2}$ as illustrated in Fig.~\ref{footprint}(b).

\begin{figure}
  \begin{center}
    \includegraphics[scale=0.43]{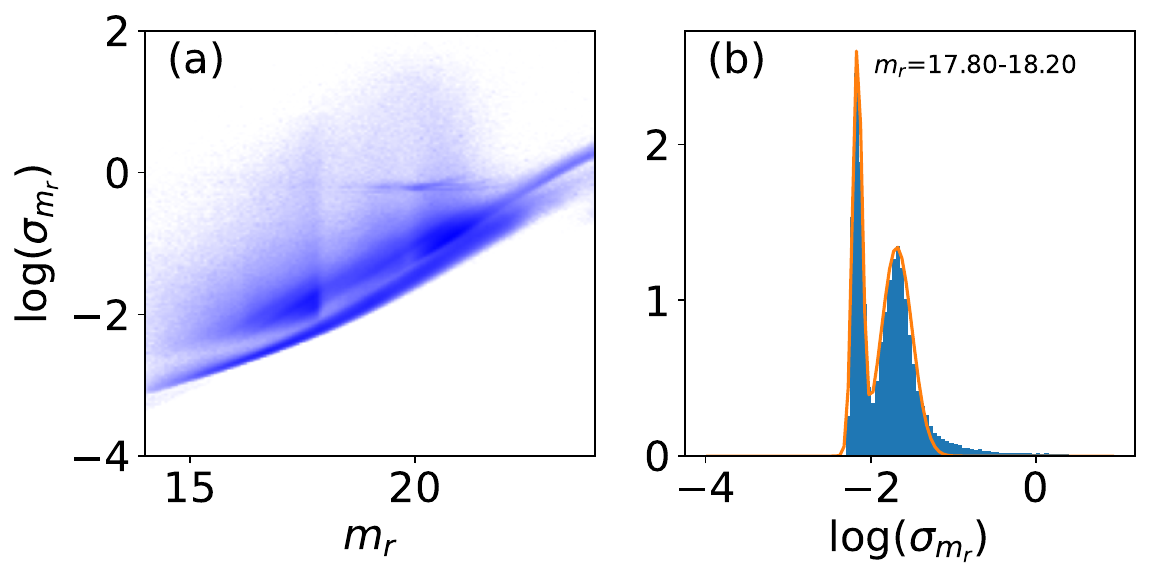}
    \includegraphics[scale=0.43]{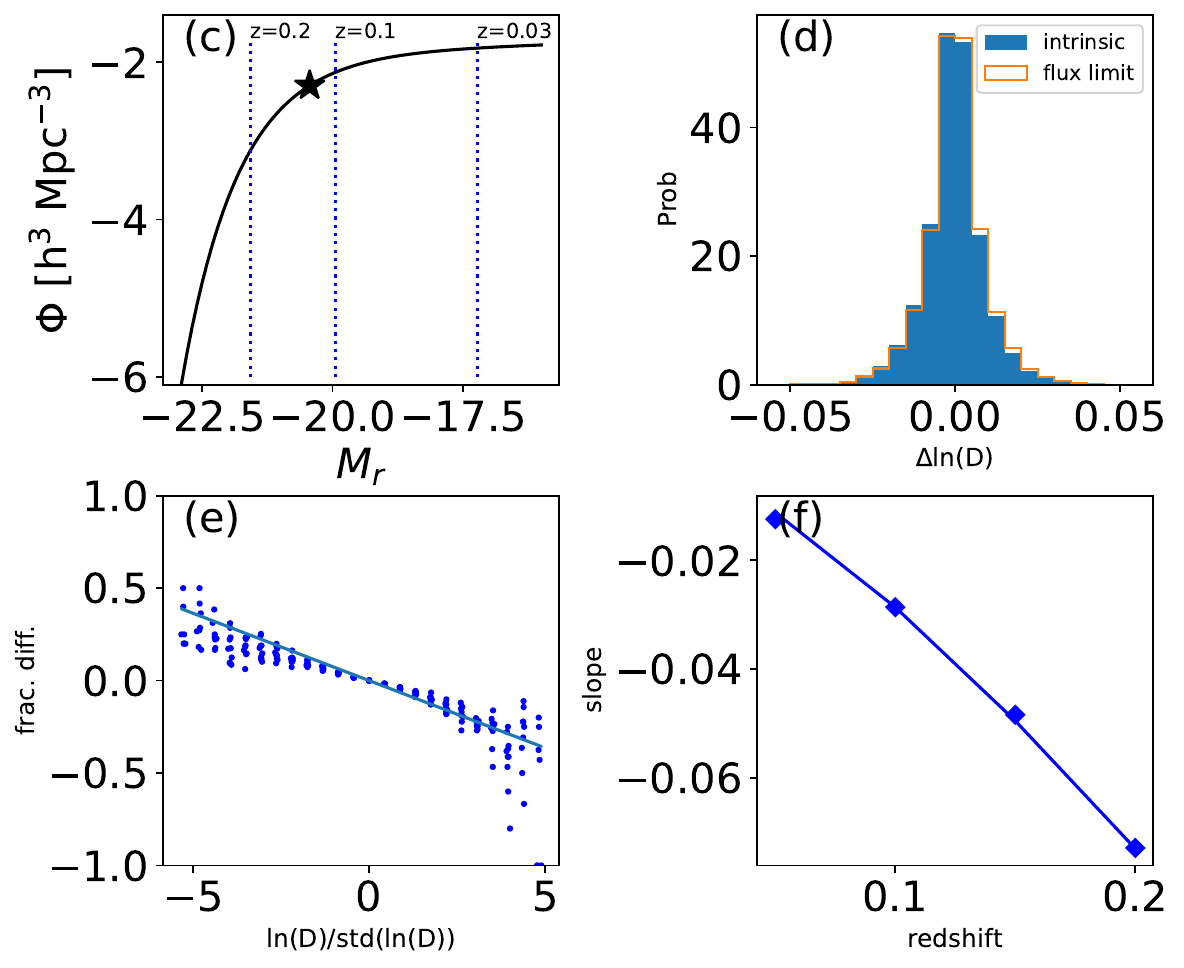}
    \caption{\label{flux_limit_corr}     {\bf (a),}    the
        $r$-band magnitude  errors versus  the apparent  magnitude for
        the entire  SDSS photometric catalog. {\bf (b),}  the distribution of
        $r$-band  magnitude  errors  in  one  magnitude  bin, along  with  the
        best-fitted double Gaussian profiles. {\bf (c),} the $r$-band luminosity
        function  with the  characteristic brightness  labeled by  a
        star symbol.  Three dotted lines mark  the absolute luminosity
        of the magnitude cut ($m_{r, \rm limit}$=17.6) at three redshifts.
        {\bf (d),} the  distribution of the difference  between the intrinsic
        distance  and  observed distance at $z$=0.2, which  are  derived from  the
        intrinsic magnitude and observed magnitude, respectively.  The
        filled histogram  represents the  one with  intrinsic apparent
        magnitude  below  $m_{r,  \rm limit}$,  while  open  one  for  the
        observed apparent magnitude below $m_{r, \rm limit}$. {\bf (e),} the fractional difference between intrinsic and
        observed distribution (the denominator is the observed distribution) as a function
        of the offset from the mean in terms of standard deviation of the distribution at $z$=0.2. The result contains
        ten simulations. The solid line is the best-fitted linear function with zero intercept. {\bf (f),} the slope of above
    linear function as a function of redshift, where the solid line is the best-fitted power law function. } 
  \end{center}
\end{figure}

\subsection{The velocity zero point}

In  the above  approach to  calculate PV,  we use  the
Gaussian mean of  the distance distribution within  a small redshift
bin as the velocity zero point.  This assumption is correct as along
as the large scale structure within the redshift bin is not coherent
so that  no bulk motion  exists. The best way  to assure this  is an
all-sky survey  that covers  both northern and  southern hemisphere.
However the SDSS  is not large enough especially  for galaxies below
$z$=0.1. Fortunately,  the SDSS  MGS have additional  coverage along
the  equatorial  plane besides  a  continuous  area in  the  northern
hemisphere.  As a result, the  Gaussian mean is measured only using
galaxies in two  great circles  of the  sky sphere,  i.e., those  along the
equatorial  plane plus  a  stripe along  longitude  as enclosed  by the
dashed boxes in Figure~\ref{footprint} (a). We demonstrate this with mock data
in \S~\ref{zero_point_mock}.

\subsection{The correction for the flux limit}

The  SDSS MGS is  a flux-limited sample. Due  to the
  luminosity function, sources that  are intrinsically faint and below
  the flux  limit have a higher  likelihood of scattering to  be above
  the limit, compared to sources  that are intrinsically bright, which
  have a  lower chance of falling  below the limit. As  a result, some
  fraction  of  sources  that  are  above the  flux  limit  should  be
  intrinsically  faint,  thus  overestimating its  stellar  mass  and
  underestimating its distance.

The log distribution of $r$-band magnitude errors as
  a  function of  the magnitude  is shown  for the  entire photometric
  catalog  from $m_{r}$=14  to 24  in Figure~\ref{flux_limit_corr}(a).
  We  fit  a double  Gaussian  function  to  the log  distribution  of
  magnitude errors in magnitude  bins with $\Delta$$m_{r}$=0.4 mag. An
  example      of     the      fitting      is     illustrated      in
  Figure~\ref{flux_limit_corr}(b) for $m_{r}$=17.8-18.2.  The SDSS MGS
  sample for the large scale structure has a flux limit around $m_{\rm
    r}$=17.6. Below the redshift of  0.1, this flux limit still probes
  those   below    the   characteristic   magnitude   as    shown   in
  Figure~\ref{flux_limit_corr}(c).

We    randomly   draw    3$\times$10$^{6}$   absolute
  magnitudes from the  luminosity function for a  given redshift.  The
  distribution ranges  from $M_{\rm r}$=$-$24 to  $M_{\rm r}$=23.25-DM,
  where DM  is the  distance module, with  23.25 corresponding  to the
  SDSS image 3-$\sigma$ depth.  Next, we assign the flux error to each
  absolute  magnitude  following Figure~\ref{flux_limit_corr}(a),  and
  calculate the difference in  the magnitude (${\Delta}M_{r}$) between
  the intrinsic and observed ones.  Since the mass-to-light ratio is a
  function of color but SDSS MGS is a purely flux-limited sample, this
  magnitude difference  corresponds to  a difference in  distance such
  that     ${\Delta}{\rm     ln}{D_{\rm     c}}$     =     $\frac{{\rm
      ln}(10)}{2.5}$$\frac{\beta}{(2\beta-0.25)}$${\Delta}M_{r}$.

We  then  obtain  the ${\Delta}{\rm  ln}{D_{\rm  c}}$
  distribution for  two separate samples: one  with intrinsic apparent
  magnitudes  below the  SDSS  MGS  bright limit  and  the other  with
  observed  apparent   magnitudes  below   the  limit,  as   shown  in
  Figure~\ref{flux_limit_corr}(d) for $z$=0.2.  The difference between
  these  two distributions  arises from  the fact  that the  latter is
  affected by the flux limit cut. 

In   Figure~\ref{flux_limit_corr}(e),  we   show  the
  fractional difference (the denominator is the observed distribution)
  in  individual  bins  of ${\Delta}{\rm  ln}{D_{\rm  c}}$/  std(${\rm
    ln}{D_{\rm c}}$).  Here, std(${\rm ln}{D_{\rm c}}$) represents the
  standard deviation of ${\Delta}{\rm  ln}{D_{\rm c}}$, and the offset
  is expressed in terms of standard deviations.  This approach ensures
  that  the above  trend remains  unchanged if  ${\rm ln}{D_{\rm  c}}$
  experiences increasing Gaussian  errors as in the  real case.  After
  conducting  10 simulations  for a  given redshift,  we fit  a linear
  function to the above trend with zero intercept. We obtain the slope
  of the trend for  $z$=0.05, 0.1, 0.15 and 0.2, and  then fit a power
  law  to interpolate  at any  given  redshift.  To  correct the  bias
  introduced by the flux limit for  each redshift bin in real case, we
  select  a portion  of galaxies  according  to its  offset from  the
  Gaussian mean and the linear function at given redshift, and reverse
  its sign of the offset.

\section{Method to calculate the momentum power spectrum}

\subsection{The methodology}
The line-of-sight mass-weighted momentum field is described as
\begin{equation}
  F^{P}(\mathbfit{r}) = (1+\delta(\mathbfit{r}))v_{\rm p}(\mathbfit{r}),
\end{equation}
where $\delta(\mathbfit{r})$ is the galaxy over-density at the location vector
$\mathbfit{r}$. For a catalog with discrete galaxies, it can be written as \citep{Feldman94, Hand17, Howlett19}:
\begin{equation}
F^{P}(\mathbfit{r}) = \frac{w(\mathbfit{r})n_{\rm g}(\mathbfit{r})v_{\rm p}(\mathbfit{r})}{A},
\end{equation}
where $w(\mathbfit{r})$ is the galaxy total weight that is $\geq$ 1,  $n_{\rm g}(\mathbfit{r})$ is the observed galaxy
number density, and $A$ is the normalization given in Equation~\ref{eqn_normalization} below.

The momentum power spectrum $P^{P}(\mathbfit{k})$=${\langle}$$F^{P}(\mathbfit{k})F^{P}(\mathbfit{k}')$${\rangle}$, the Fourier analog of the two-point correlation function
 ${\langle}F^{P}(\mathbfit{r}) F^{P}(\mathbfit{r}'){\rangle}$ where
 $\mathbfit{k}$=$\frac{2{\pi}(\hat{\mathbfit{r}}-\hat{\mathbfit{r}}')}{|\mathbfit{r}-\mathbfit{r}'|}$
\footnote{The hat symbol denotes the unit vector.}, can expand through Legendre polynomials:

\begin{equation}
  P^{P}(\mathbfit{k}) = {\sum_{\ell}} P^{P}_{\ell}(k)L_{\ell}(\mu),
\end{equation}
where $\mu = \hat{\mathbfit{k}}\cdot\hat{\mathbfit{r}}_{h}$ and
$\hat{\mathbfit{r}}_{h}=(\hat{\mathbfit{r}}+\hat{\mathbfit{r}}')/2$.
The estimator for the $P^{P}_{\ell}(k)$ is given by \citep{Feldman94, Hand17}:
\begin{equation}
\begin{aligned}
  P^{P}_{\ell}(k) & = \frac{2\ell+1}{A^{2}}\int\frac{d\Omega_{k}}{4\pi}
   \biggl[ {\int}d\mathbf{r}{\int}d\mathbf{r}'F^{P}(\mathbfit{r}) F^{P}(\mathbfit{r}') \\
  &  e^{i\mathbfit{k}\cdot(\mathbfit{r}-\mathbfit{r}')}L_{\ell}(\hat{\mathbfit{k}}\cdot
   \hat{\mathbfit{r}}_{h}) - P^{\rm P, noise}_{\ell}(\mathbfit{k}) \biggl], \\
\end{aligned}
\end{equation}
where $\Omega_{k}$ is the solid angle in Fourier space.
By assuming the local plane-parallel approximation of
($\hat{\mathbfit{k}}\cdot \hat{\mathbfit{r}}_{h}$)$\approx$($\hat{\mathbfit{k}}\cdot\hat{\mathbfit{r}}$)$\approx$($\hat{\mathbfit{k}}\cdot\hat{\mathbfit{r}}'$), the integral over $F^{P}(\mathbfit{r})$
and  $F^{P}(\mathbfit{r}')$ in the above equation can be separated into independent integrals
\citep{Yamamoto06}. By further decomposing the $\hat{\mathbfit{k}}{\cdot}\hat{\mathbfit{r}}$
into a product of spherical harmonics, the calculation of the above equation can be done
with a small number of 
Fast Fourier Transforms  (FFTs), which significantly saves the
computing time \citep{Hand17}.  In this study, we are only interested
in $\ell$=0, i.e. the angle-averaged momentum power spectrum $P^{P}_{\ell=0}(k)$, so that the local
plane-parallel is exactly valid.


\begin{figure*}
  \begin{center}
    \includegraphics[scale=0.5]{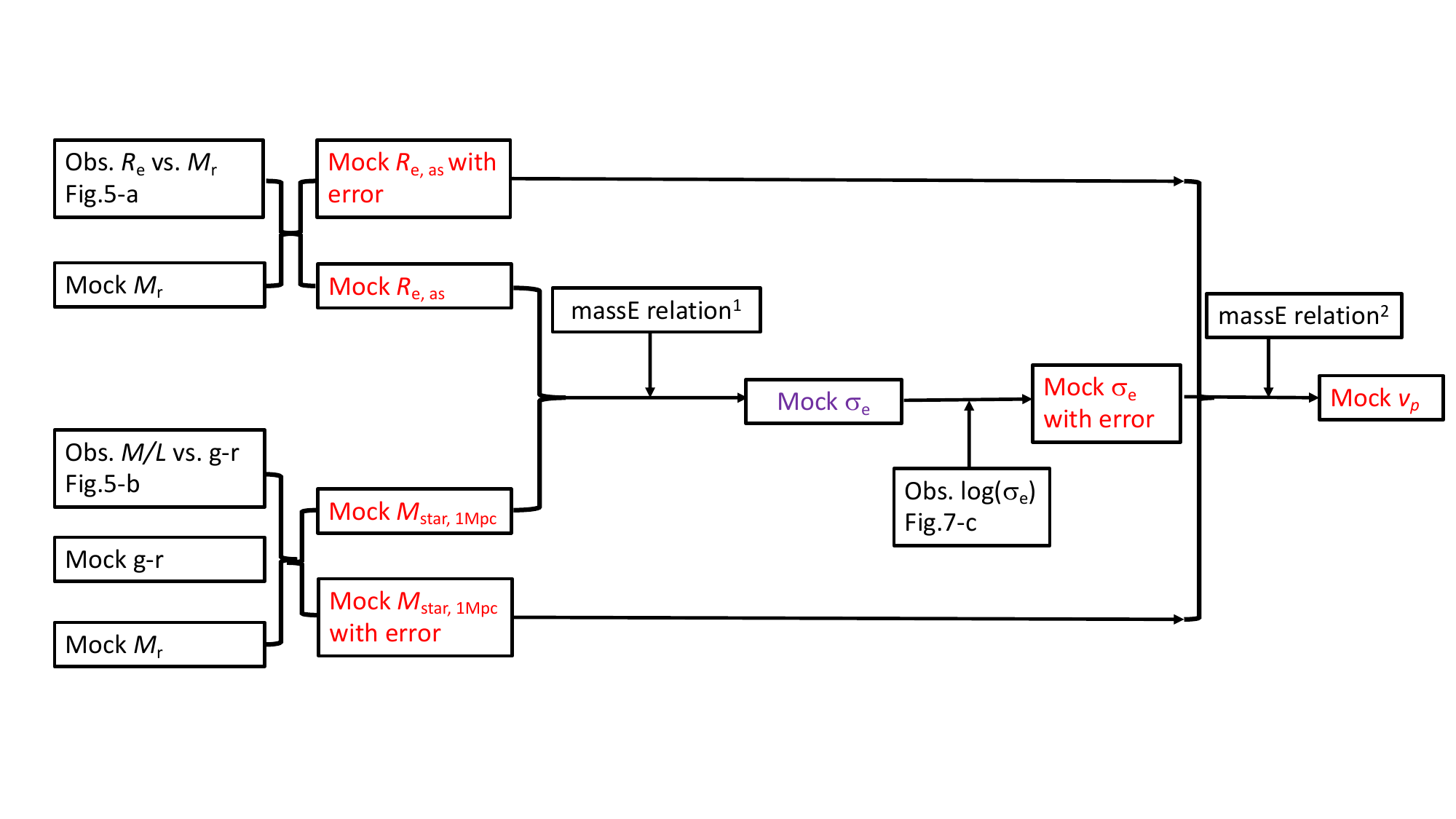}
    \vspace{-1.5cm}
    \caption{\label{mock_outline} The flow chart to mimic  the observed
      PV for Uchuu-SDSS mock galaxies. The first massE relation is the intrinsic one, and the second
    relation is used to measure PV. We test the cases if they are different in \S~\ref{sec_comp_vpec_mock}.}
\end{center}
\end{figure*}

\begin{figure*}
  \begin{center}
    \includegraphics[scale=0.5]{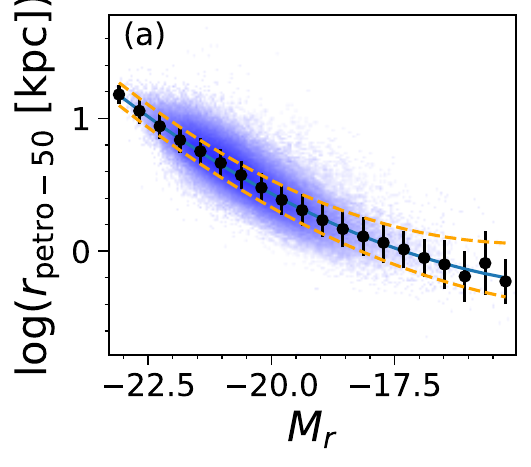}
    \includegraphics[scale=0.5]{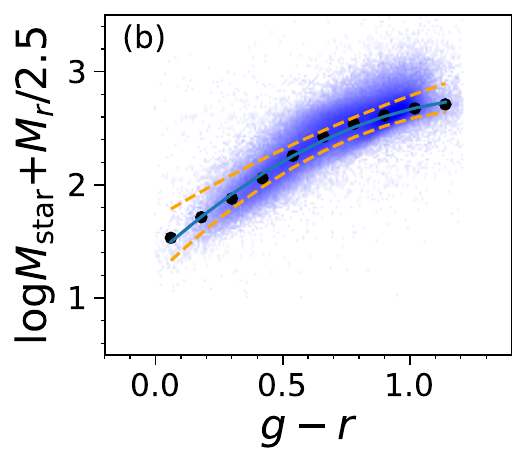}
    \includegraphics[scale=0.5]{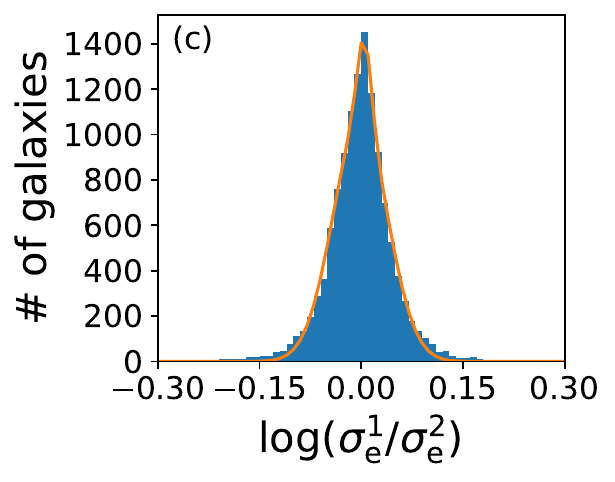}
    \caption{\label{relationships_SDSS} {\bf (a)}, the observed physical
      size vs. absolute $r$-band magnitude of the SDSS MGS elliptical galaxies. The solid
      line is a second-order polynomial fit to the median trend, and two dashed lines represent the standard
      deviation of the trends.
      {\bf (b)}, the mass-to-light ratio in $r$-band vs. $g$-$r$ color of all SDSS MGS galaxies.
      The solid and dashed lines are the second-order polynomial fit to the median and standard deviations of
      trends, respectively. {\bf (c)}, the distribution of the velocity dispersion errors based on duplicate measurements of about 10$^{4}$ objects
      in the MPA-JHU catalog.}

\end{center}
\end{figure*}

\subsection{The implementation of the methodology}

The python code \texttt{nbodykit}  \citep{Hand18} implements the above
calculation  for  the  density  power spectrum.  We  make  some  minor
modification   of   their   python   codes   \texttt{FKPCatalog}   and
\texttt{ConvolvedFFTPower} to  produce \texttt{FKPMomentumCatalog} and
\texttt{MomentumConvolvedFFTPower}  for  the momentum  power  spectrum
calculation. Here we list the key inputs in order to run the code:

(1) $w_{\rm  mpa-jhu}$ -- the  completeness weight due to  the failure
match of the  LSS catalog with the MPA-JHU catalog:  the reciprocal of
\completenessMatchMPAJHU\; as derived in Table~\ref{tab_sel_sample} is
taken as $w_{\rm mpa-jhu}$ for  all galaxies.

(2) $w_{v_{\rm  p}}$ --  the completeness weight  for the  $v_{\rm p}$
measurement that is due to the failure in measurements of $\sigma_{\rm
  e}$: we divide the RA, DEC and $z$-range of our sample into 10 bins,
respectively. In each 3-D cell,  the completeness is calculated as the
ratio of  the number  of objects  with available  $v_{\rm p}$  and the
total number. The  completeness is found to be almost  a constant with
the  median value  of \completenessVp\;  and a  standard deviation  of
\completenessVpSD.  As  a  result  we  assign  $w_{v_{\rm  p}}$  as  a
reciprocal of the median completeness for all galaxies.

(3) $w_{\rm sel}$ -- the weight for the selection function of the SDSS
main galaxy LSS sample.  This number is available in
the catalog, which is 1/\texttt{fgotten}.

(4) The observed number density $\bar{n}(z)$ as a function of redshift: this function is shown in
Figure~\ref{nofz_v2ofz}(a). We interpolate the function to get the number density at the redshift of
each galaxy.

(5) The $\langle{v_{\rm p}^{2}}\rangle_{z}$ as a function of redshift: this function is shown
in Fig.~\ref{nofz_v2ofz}(b). Similarly we interpolate it to obtain the
velocity variance at the redshift of each galaxy.

(6) The FKP weight 
\begin{equation}
w_{\rm FKP} = \frac{1}{\langle{v_{\rm p}^{2}}\rangle_{z} + \bar{n}(z)P^{\rm P}},
\end{equation}
where $P^{\rm P}$ is set to be $10^{9}$ \unitP, which is roughly the momentum
power spectrum at our median spatial scale of 0.1 $h$/Mpc.

(7) The  range of  spatial scales  as bracketed  by $k_{\rm  min}$ and
$k_{\rm  max}$:  $k_{\rm  min}$  is crucial  for  the  measurement  of
structure growth that is sensitive  to large spatial scales. We set it to be 0.02  $h$/Mpc.
  Our test with mock data shows that
the momentum power spectrum is recovered well above  $k_{\rm  min}$. 
For $k_{\rm max}$, we test several values and found that
the growth rate is insensitive to it.

(8) With the combined completeness weight
$w_{\rm complete}$=$w_{\rm mpa-jhu}$$w_{v_{\rm p}}$$w_{\rm sel}$ and FKP weight
$w_{\rm FKP}$, we ran \texttt{FKPMomentumCatalog} and \texttt{MomentumConvolvedFFTPower} implemented in \texttt{nbodykit} to get the momentum power spectrum which is
further subtracted by the shot-noise contribution:
\begin{equation}
  P^{\rm P, noise}_{\ell=0} = \frac{ \sum \langle{v_{\rm p}^{2}}\rangle_{z} w_{\rm FKP}^{2} w_{\rm complete}^{2} }{A},
\end{equation}
where $A$ is the normalization as given by
\begin{equation}\label{eqn_normalization}
  A = {\sum w_{\rm FKP}^{2} \bar{n}(z) w_{\rm complete}}.
\end{equation}
Given $w_{\rm tot}$=$w_{\rm complete}$$w_{\rm FKP}$, the effective redshift of the sample is
given by
\begin{equation}
  z_{\rm eff} = \frac{\sum w_{\mathrm{tot},i}w_{\mathrm{tot},j}(z_{\mathrm{obs},i}+z_{\mathrm{obs},j})}
  {2\sum w_{\mathrm{tot},i}w_{\mathrm{tot},j}},
\end{equation}
where $i$ and $j$ are members of all galaxy pairs in the sample. For our whole sample,
 $z_{\rm eff}$ is found to be \zeff.

\section{Mocks of momentum power spectra}\label{SecValidVpec}

\subsection{Mock galaxy catalogs and true momentum power spectrum}\label{sec_mock_true}

We use  32 Uchuu-SDSS mock  galaxy catalogs that are  constructed from
the  Uchuu 2.1  trillion N-body  simulation \citep{Dong-Paez22}.   The
catalogs match the  depth, footprint, selection function etc of the SDSS MGS in the North Cap region.
To construct the mock catalog, the $r$-band absolute magnitude $M_{r}$
was  assigned to  a  halo through  subhalo  abundance matching  (SHAM)
techniques.  The  $g$-$r$ color was further  assigned randomly through
the  probability distribution  of the  color-magnitude diagram.   With
$M_{r}$, $g$-$r$  color and  observed redshift, k-d  tree was  used to
find  the  closest  real  galaxy  for  each  mock  galaxy,  after  which
properties of real  galaxies were attached to mock  galaxies.  For our
use, we  need one additional  parameter -- \texttt{FRACDEV},  which is
obtained with k-d tree match also  based on $M_{r}$, $g$-$r$ color and
observed redshift.

\begin{figure*}
  \begin{center}
    \includegraphics[scale=0.55]{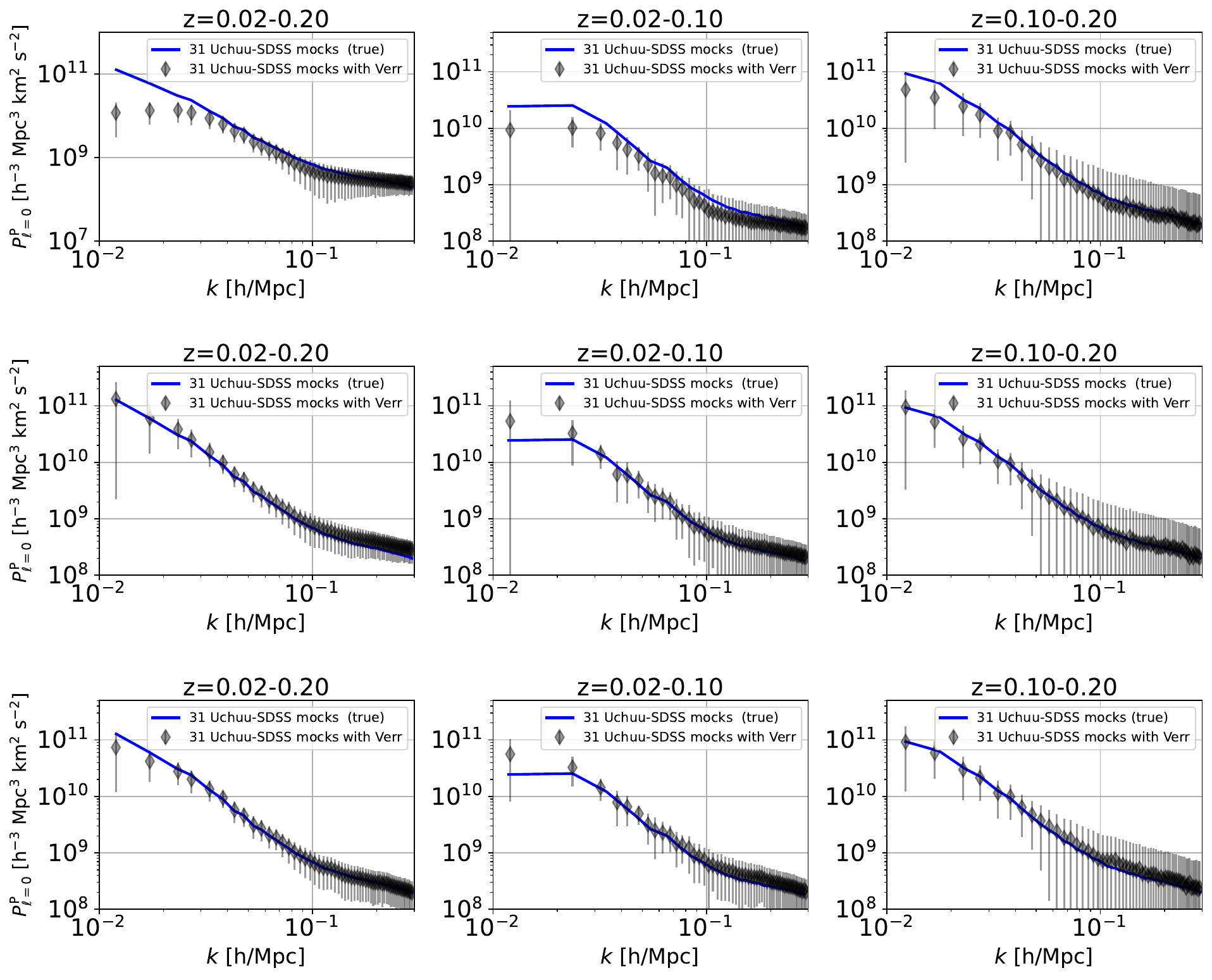}
    \caption{\label{mom_power_spec_mock_zeropoint} The first row: for each mock (mock 0 to mock 30),
      the zero velocity point (Gaussian mean) is determined from galaxies in that particular mock;
      the second row: the zero velocity point is determined
      from  galaxies in that particular mock plus all galaxies in mock 31; the bottom row:
      for each mock, the zero velocity is determined from part of galaxies in mock 31 and part of galaxies in that particular mock to mimick galaxies in yellow dashed boxes in Figure~\ref{footprint}. For details,
      see \S~\ref{zero_point_mock}. From left to right, the momentum power spectra are presented
      for three redshift ranges. The solid line is median of the intrinsic spectra of 31 mocks
      (mock 0 to mock 30), and symbols are the median of 31 mocks calculated with the massE-based PVs. }
\end{center}
\end{figure*}

We define elliptical galaxies of  each mock catalog following the same
criteria  as  the observed  one:  (1)  \texttt{FRACDEV} $\geq$ 0.8 and  (2) 0.02
$<$ $z_{\rm obs}$ $<$ 0.2.   With true peculiar velocities obtained through
Equation~\ref{eqn_zobs_zcos}, the true momentum power spectrum is then
calculated    using
\texttt{nbodykit}. The  median  of  intrinsic momentum
power spectra of  mocks is presented 
  as solid lines in Figure~\ref{mom_power_spec_mock_zeropoint}.

\subsection{Mimicking PV measurements}\label{sec_sim_pec_mock}

To produce the PV for the mock galaxy catalog,
we carry out the following simulation as outlined in Figure~\ref{mock_outline}:

(1)  The apparent  half-light radius  in arcsec  ($R_{\rm e,  as}^{\rm
  mock}$) for  each mock galaxy:  we first plot the  observed physical
half-light radius as a function  of $M_{r}$ of the SDSS MGS-elliptical
sample   in  Figure~\ref{relationships_SDSS}(a).    Here  we   include
ellipticals with  $z_{\rm obs}$=0.01-0.2 to extend  the dynamic range.
As shown  in the figure, we  fit the median and standard deviations of half-light radii
at given $M_{r}$ with second-order  polynomial functions, respectively.
We then assign  the intrinsic physical  size to  each mock galaxy based on the above median trend along with  its mock
$M_{r}$. The associated errors are assigned based on standard deviations of  trends
(dashed lines in Figure~\ref{mock_outline} (a)). The mock physical
size  is then further  converted to  the angular  size $R_{\rm  e, as}^{\rm
  mock}$ based on the mock true angular diameter distance.

(2) The  stellar mass  normalized at 1  Mpc ($M_{\rm  star, 1Mpc}^{\rm
  mock}$)  for   each  mock  galaxy:   we  first  plot   the  observed
mass-to-light ratio in terms of log$M_{\rm star}$+$M_{\rm r}$/2.5 as a
function   of   $g$-$r$    color   based   on   the    SDSS   MGS   in
Figure~\ref{relationships_SDSS}(b).   Here  we  include all  types  of
galaxies in order to extend  the dynamic range. Again, second-order polynomial
functions  are  fitted to  the  median and standard deviations of  the
relationship, respectively.   We then  assign an intrinsic stellar mass  to each
mock galaxy based on its mock $g$-$r$ color and $M_{\rm r}$ using the median trend, while the observed
errors are assigned based on the fitting to the standard deviations of  the
relationship (dashed lines in Figure~\ref{mock_outline} (b)).  The  stellar mass is then  converted to
that at 1 Mpc ($M_{\rm star, 1Mpc}^{\rm mock}$) based on the mock true
luminosity distance.

(3)  The velocity  dispersion ($\sigma_{\rm  e}^{\rm mock}$)  for each
mock  galaxy:   we  first  use   Equation~\ref{eqn_distance_ruler}  to
calculate  an  intrinsic  velocity   dispersion  from  the intrinsic
$M_{\rm  star,  1Mpc}^{\rm  mock}$ and  $R_{\rm  e, as}^{\rm  mock}$ along
with the  mock  true distance, for which the slope $\beta$ of the massE
relationship is fixed to the best-fit
value in Equation~\ref{eqn_beta}.
About 10$^{4}$ objects of the PV sample have duplicate measurements of the
velocity dispersion in the MPA-JHU catalog.
As shown in Figure~\ref{mock_outline} (c), the distribution of the
velocity dispersion errors can be described by two Gaussian profiles. We thus assign each galaxies
following the error distribution in the figure.

(4) The mock distance for each  mock galaxy: based on the mock $R_{\rm
  e, as}^{\rm mock}$, $M_{\rm star, 1Mpc}^{\rm mock}$ and $\sigma_{\rm
  e}^{\rm mock}$ with  errors, we calculate the  mock distance through
Equation~\ref{eqn_distance_ruler}. Here we perturb the $\beta$ and the
corresponding $D_{0}$ with their errors, respectively. 

(5) The  mock PV for  each mock galaxy: we  then follow
the same procedure as the observed one to obtain the PV
for each mock galaxy.   For each  of 31 mocks from mock-0 to mock-30,  we  run  60 simulations so that in total
we have 1860 simulations for  each
observed  set  of  PV  measurement. Note that the last mock (mock-31) is
  used to test the zero velocity point.   Each  simulation
includes  random perturbation  of  the intrinsic  $R_{\rm e,  as}^{\rm
  mock}$,  $M_{\rm star,  1Mpc}^{\rm mock}$  and $\sigma_{\rm  e}^{\rm
  mock}$  with  their error  distributions,  and  one perturbation  of
$\beta$  and  $D_{0}$,  respectively.

By construction, the mock catalog does not have failure measurements, so we do not
remove any outliers. We confirm that the velocity errors produced by mocks are similar to
  the observed ones at different redshift.

\subsection{The velocity zero point}\label{zero_point_mock}

In this section,  we examine the impact of the  velocity zero point on
the momentum power spectrum. In our method to measure the PV, the zero
velocity  represents the  Gaussian mean  of the  distance distribution
within a  narrow redshift bin.  We  first present the case  where each
mock  uses its  own distance  distribution to  calculate the  Gaussian
mean. As a result, the zero point is derived from galaxies in an about
7000 deg$^{2}$ area in the northern hemisphere.  As shown in the first
row of Figure~\ref{mom_power_spec_mock_zeropoint}, in the low redshift
range  of  $\Delta$z=0.02-0.10  the derived  momentum  power  spectrum is
systematically  lower  than  the  intrinsic  one,  while  the  one  in
$\Delta$z=0.10-0.20  overall  recovers  the intrinsic  one  very  well
except for  small k that is  below 0.02 $h$/Mpc. This  demonstrates that
because of  the limited  volume at  low redshift  a small  bulk motion
still exists and affects our methodology.

The Uchuu-SDSS provides 32 independent  mocks to simulate the SDSS MGS
in the  northern hemisphere. In the  second case, we use  the last
mock (mock-31) to  emulate a SDSS-MGS-like  survey in  the southern
hemisphere. For each of the remaining mocks, we combine their galaxies
with those  from mock 31  to measure the  Gaussian mean. This  mean is
then used to derive PV for the galaxies in that particular mock.  Note
that  galaxies  from mock  31  are  solely  used for  determining  the
Gaussian mean and are not used  in the calculation of the momentum
power    spectra.     As    shown    in    the    middle    panel    of
Figure~\ref{mom_power_spec_mock_zeropoint},  the   intrinsic  momentum
power   spectra  for   three  different   redshift  ranges   are  well
recovered. This indicates, even at redshifts below $z$=0.1, the effect
of  the   velocity zero point  on  the  momentum  power  spectrum  is
negligible if large enough survey is conducted.

Unfortunately, the SDSS MGS lacks a comparable survey in the southern
hemisphere.  But  it  does   contain  galaxies  along  the  equatorial
plane. In the  third case, we thus combine part  of galaxies from mock
31 with part  of galaxies from each of the  remaining mocks to calculate
the Gaussian mean for that mock. For  the part from mock 31, we select
galaxies with  a declination below  5 degrees, which roughly  yields a
similar  sky area  to  the part  of  the  SDSS MGS  that  lies in  the
equatorial plane but not in  the continuous northern regions. This sky
area is represented by the two orange boxes along the equatorial plane
in Figure~\ref{footprint}.  For each of the remaining mocks, we select
their  galaxies along  the longitude  plane  that is  similar to  the
stripe in Figure~\ref{footprint}. The above  two parts of galaxies lie
on  two  perpendicular great  circles  of  the  sky sphere,  and  thus
effectively reduce  the coherent  bulk motion. As  shown in  the lower
panel of Figure~\ref{mom_power_spec_mock_zeropoint},  in both the high
and  low  redshift bins,  the  intrinsic  momentum power  spectra  are
recovered reasonably well. We thus use this strategy for both the observation data  and
mock data.

\begin{figure}
  \begin{center}
    \includegraphics[scale=0.5]{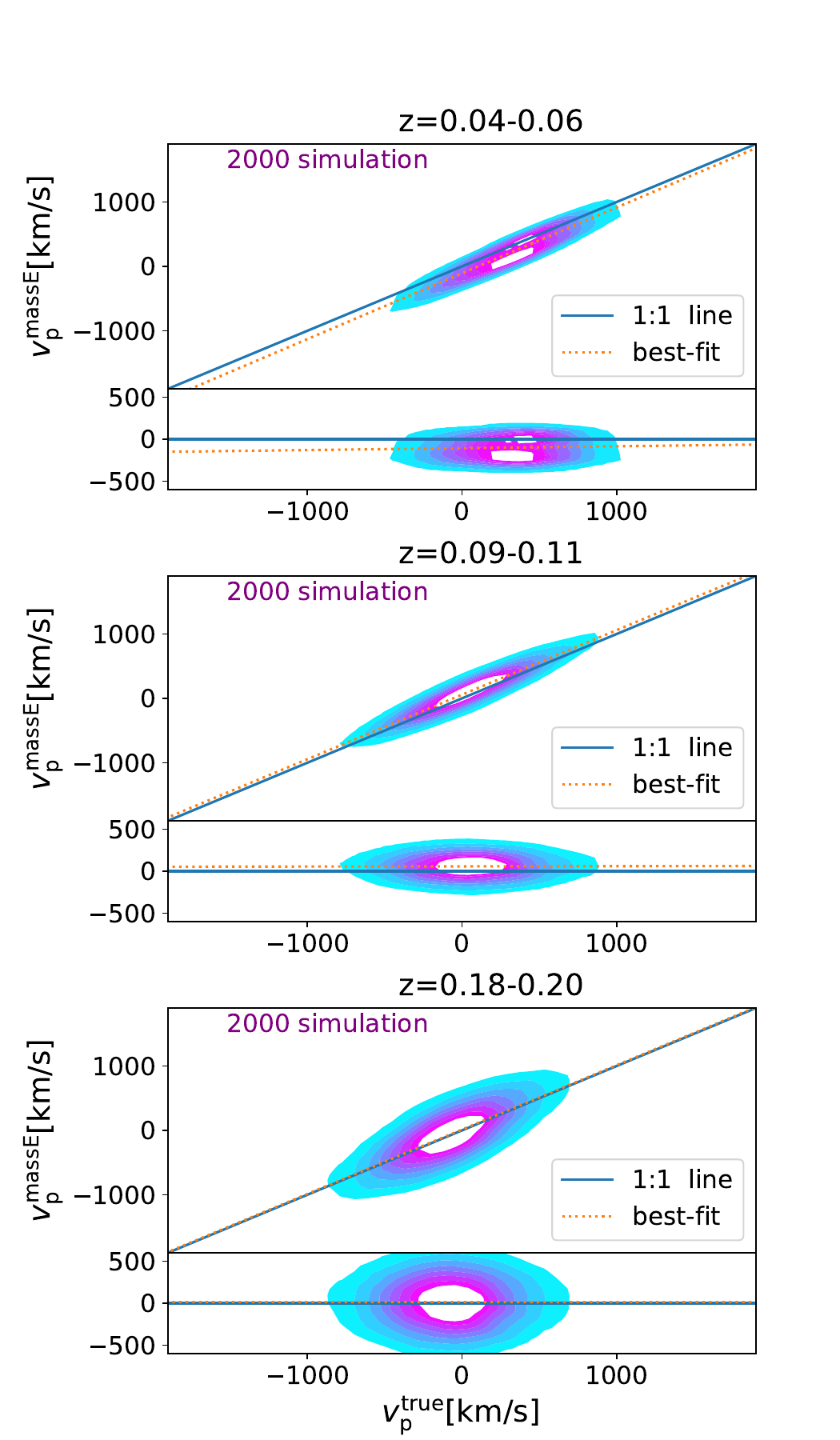}
    \caption{\label{vmock_vtrue}  The mean
        of the massE-based PV after 2000 simulations versus
        the true PVs in mocks. The lower panel for each redshift shows the
    residual between two velocities.}
\end{center}
\end{figure}

\begin{figure*}
  \begin{center}
    \includegraphics[scale=0.7]{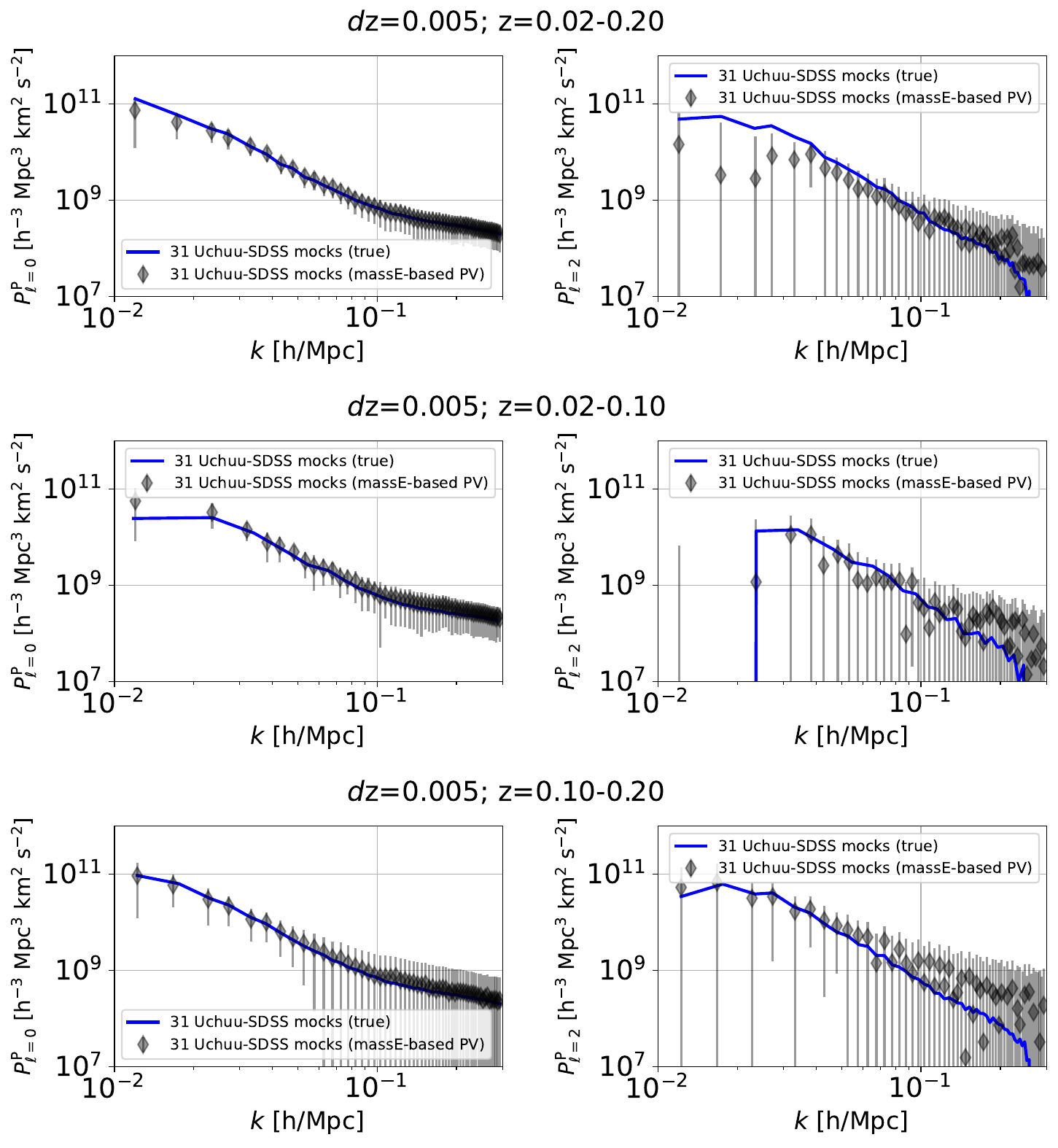}
    \caption{\label{mom_power_spec_mock_Verr_vs_noVerr}
       The median of the intrinsic momentum power spectra
      of Uchuu-SDSS mock catalogs (solid line), as compared to its 1860
      simulations of PV measurements that mimic observations
      (symbols are the median and error bars are the square root of the diagonal covariance).} 
    
    
\end{center}
\end{figure*}

\begin{figure}
  \begin{center}
    \includegraphics[scale=0.5]{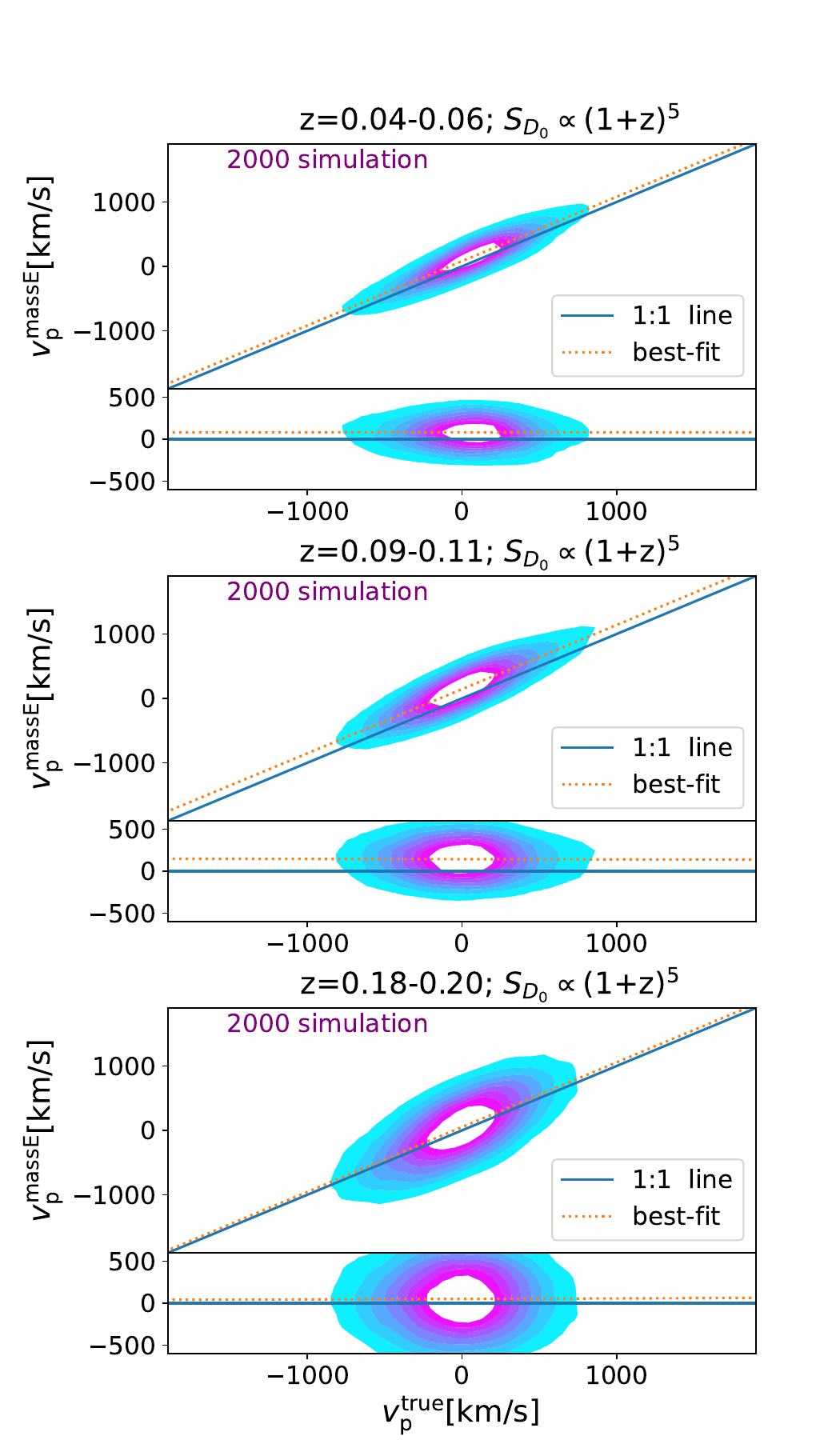}
    \caption{\label{vmock_vtrue_D0_zevl} The same as Figure~\ref{vmock_vtrue} but now the intrinsic
      massE in the mock has no redshift  evolution, while the one used to derive PV has a strong
      redshift evolution in the intercept with $S_{D_{0}}$ $\propto$ $(1+z)^{5}$.} 
\end{center}
\end{figure}

\begin{figure}
  \begin{center}
    \includegraphics[scale=0.5]{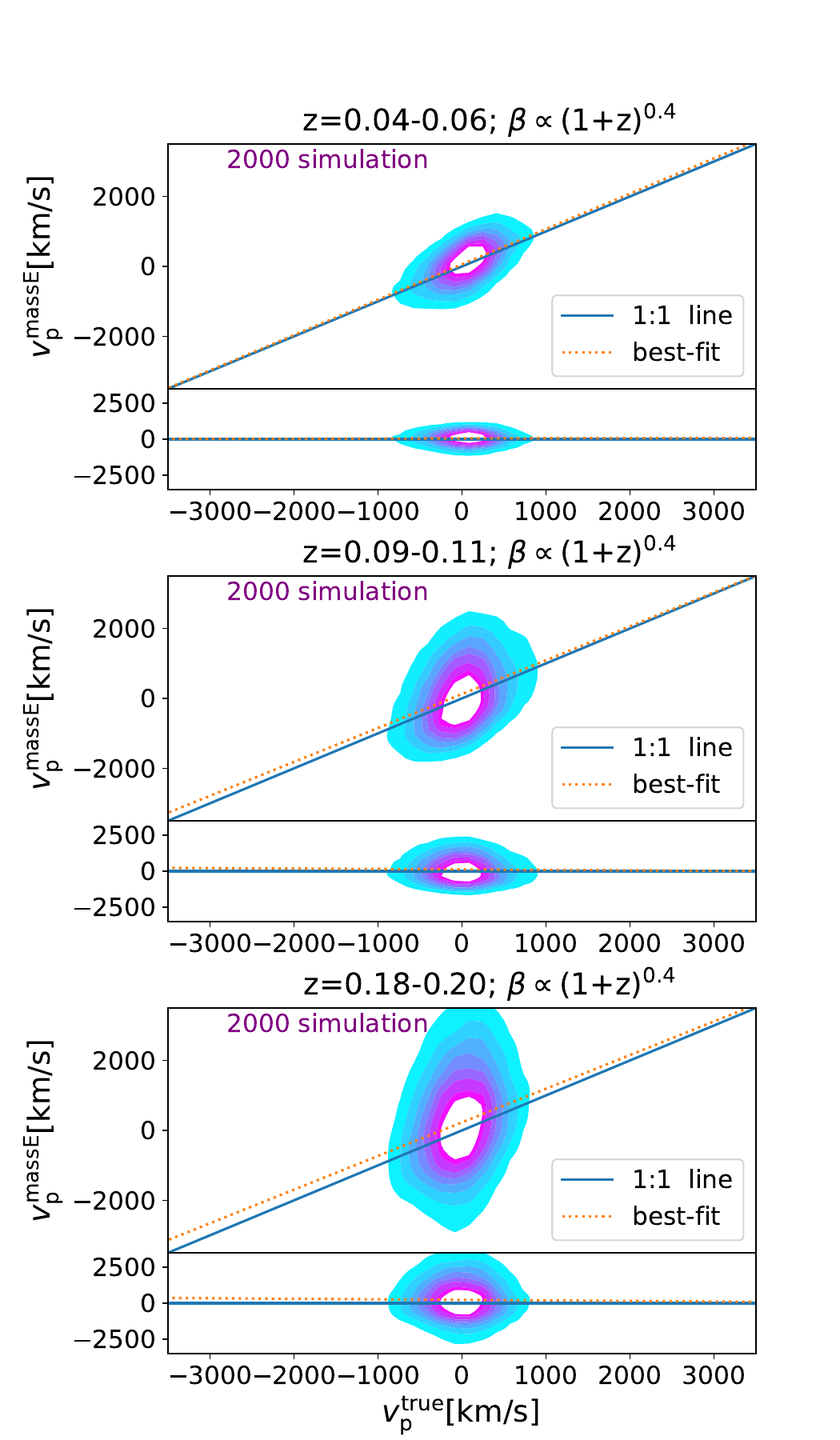}
    \caption{\label{vmock_vtrue_beta_zevl} The same as Figure~\ref{vmock_vtrue} but now the intrinsic
      massE in the mock has no redshift  evolution, while the one used to derive PV has a strong
      redshift evolution in the slope with $\beta$ $\propto$ $(1+z)^{0.4}$.} 
\end{center}
\end{figure}

\begin{figure}
  \begin{center}
    \includegraphics[scale=0.35]{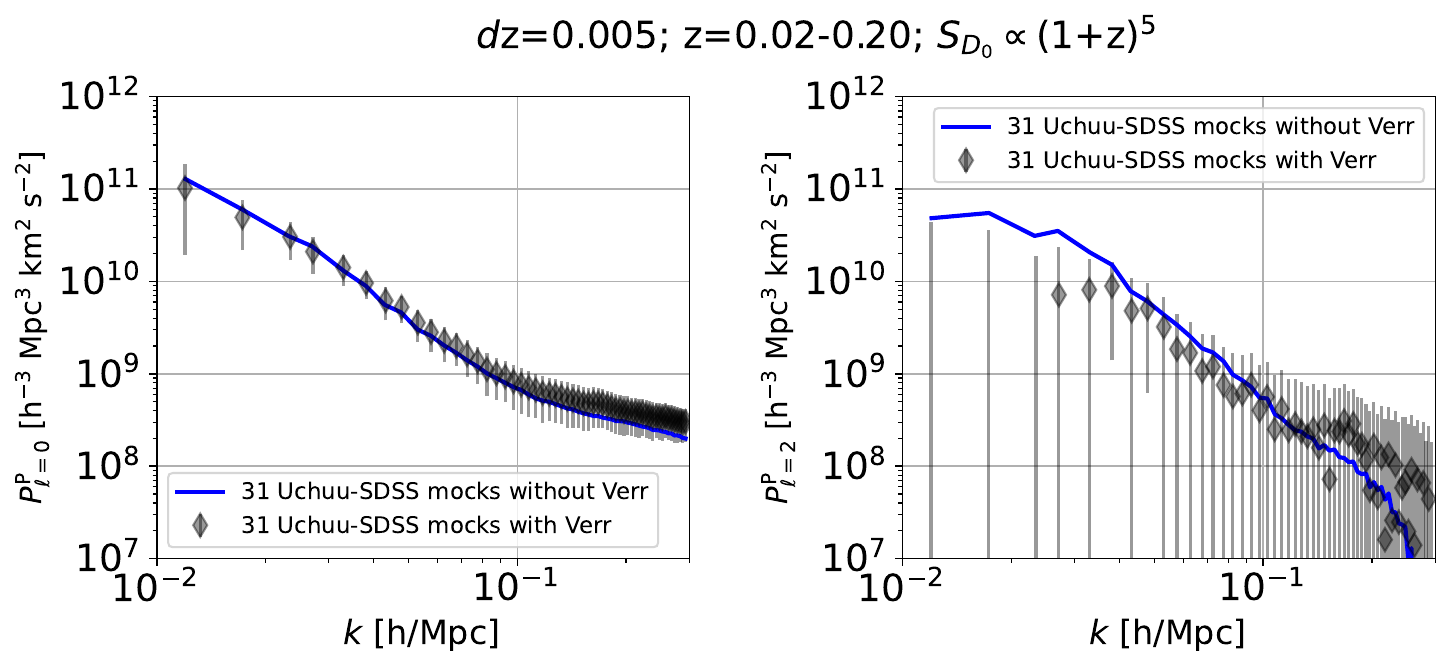}
    \includegraphics[scale=0.35]{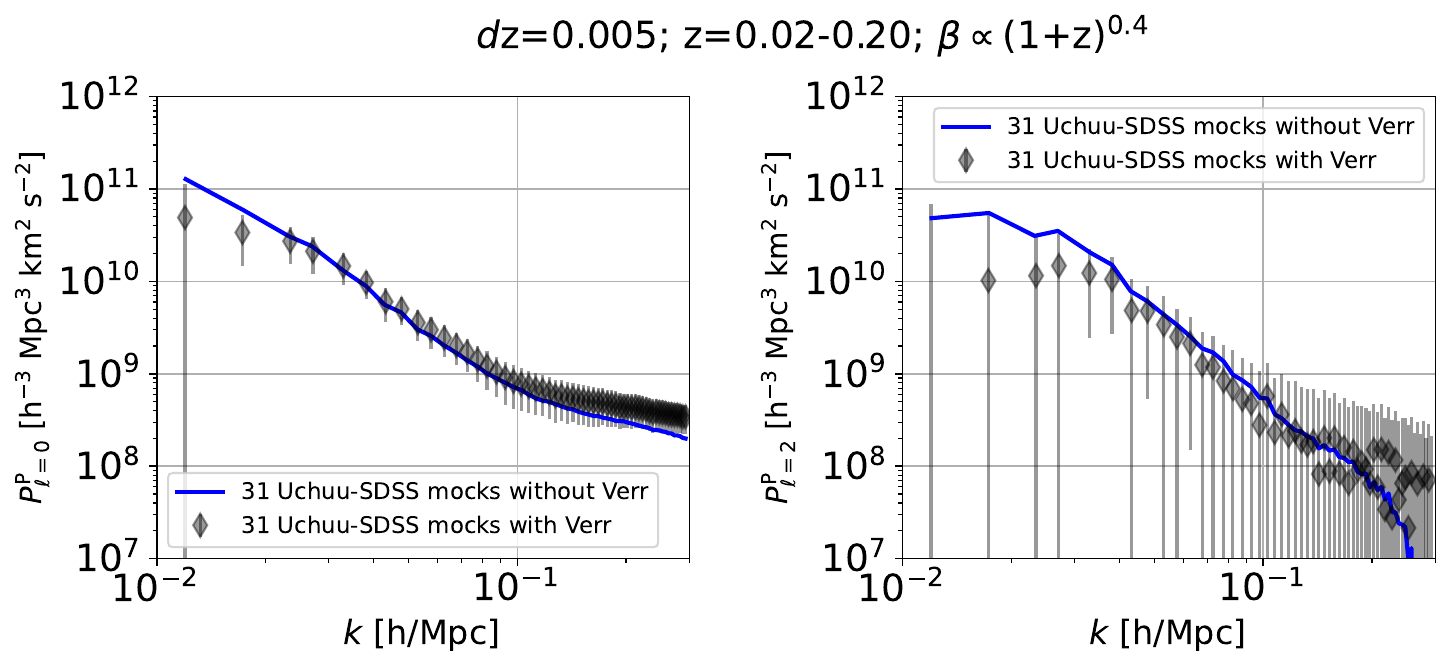}
    \caption{\label{mom_power_spec_mock_Verr_vs_noVerr_zevl} The same as Figure~\ref{mom_power_spec_mock_Verr_vs_noVerr} but only for the entire redshift range and the case that the intrinsic
      massE in the mock has no redshift  evolution while the one used to derive PV has a strong
      redshift evolution in the intercept and slope, respectively. }
\end{center}
\end{figure}

\subsection{Comparisons between massE-based measurements and true ones
in mock data-sets}\label{sec_comp_vpec_mock}

 To assess  the validity of our  methodology, we conduct two  types of
 comparison between the massE-based measurements  and the true ones in
 mock  data-sets.  The  first one  is the  PV of  individual galaxies,
 while the  second focuses on  the momentum power  spectrum, including
 both the monopole and quadrupole. Figure~\ref{vmock_vtrue} presents a
 comparison  between the  derived PV  and the  true velocity  in three
 redshift bins  for galaxies, where the  derived PV is the  mean value
 after 2000 simulations.   The two velocities in all
   three redshift bins exhibit an almost one-to-one correlation. This suggests
   that, on  a statistical level,  the massE effectively  recovers the
   true velocity.  A small  offset is observed  in the  zero velocity,
   which represents  a random error  on the zero velocity  point. This
   offset  adds  to the  shot  noise  \citep{Howlett19},
   resulting in a slight increase in the momentum power spectrum above
   0.1             $h$/Mpc,            as             seen            in
   Figure~\ref{mom_power_spec_mock_Verr_vs_noVerr}.    However,   this
   should not affect  the constraint on the growth rate,  as it is not
   sensitive to the high $k$ regime.  

 To   further    evaluate   the   velocity   field,
  Figure~\ref{mom_power_spec_mock_Verr_vs_noVerr}      presents      a
  comparison  of the  momentum power  spectra between  the massE-based
  measurements and true ones. The  solid line indicates the mean value
  from  31 mocks,  while the  symbols  represents the  mean values  of
  massE-based measurements for  31 mocks with each  mock consisting of
  60 simulations.  The left panel  of the figure displays the monopole
  of  the  momentum  power   spectrum  for  three  different  redshift
  ranges. The massE-based measurement  well recovers the true momentum
  power  spectrum except for a slight increment above
    $k\sim$ 0.1 $h$/Mpc as stated above.  Although  the survey  volume of  mock data  is not
  large  enough  to  measure  the quadrupole  of  the  momentum  power
  spectrum,  by  examining the  average  of  simulations it  can  give
  insights  into  whether  the  velocity field  is  recovered  by  the
  massE. Note  that the quadrupole  power spectrum is affected  by the
  wide angle effect  but for the SDSS MGS survey  volume it is minimal
  compared  to errors  in  the measurement  \citep[e.g.][]{Castorina18}.
  Additionally, when comparing the true  quadrupole to the one derived
  from massE, both  measurements should be affected  by the wide-angle
  effect  to the  same extent.   As shown  in the  right panel  of the
  figure, across the whole redshift range ($\Delta$$z$=0.02-0.20), the
  massE-based measurement follows the true quadrupole momentum with no
  systematic shift.   For two  redshift sub-ranges,  only at  high $k$
  values exceeding  0.2 h/Mpc, the massE-based  measurements appear to
  be higher than the true values, but note that the error is large too.

We perform  additional tests, taking  into account that  the intrinsic
massE  (used to  derive the  velocity  dispersion) does  not show  any
redshift evolution but we mistakenly incorporate redshift evolution in
the massE when calculating the PV (see Fig.~\ref{mock_outline} for two
relations).  The first case assumes  the intercept in the second massE
has    a    strong    redshift    evolution    with    a    form    of
$S_{D_{0}}$(z)=$S_{D_{0}}$(1+z)$^{5}$.        As       shown        in
Figure~\ref{vmock_vtrue_D0_zevl}, the  best fit  to PVs  of individual
galaxies   in  three   redshift  bins   still  show   good  one-to-one
correlations with  slightly increase  in errors.   The upper  panel of
Figure~\ref{mom_power_spec_mock_Verr_vs_noVerr_zevl}  shows that  both
monopole  and   quadrupole  of   the  momentum  power   spectrum  over
$\Delta{z}$=0.02-0.2 is  well recovered  too. The second  case assumes
the slope in  the second massE has  a redshift evolution in  a form of
$\beta$(z)=$\beta$(1+z)$^{0.4}$.  At $z$=0.2  the evolution results in
an approximately 20-$\sigma$ offset from  our best estimate of $\beta$
as            listed            in            Equation~\ref{eqn_beta}.
Figure~\ref{vmock_vtrue_beta_zevl}  shows that  best  fits  to PVs  of
individual  galaxies in  three redshift  bins also  exhibit one-to-one
correlations,   but   with  larger   errors.   The   lower  panel   of
Figure~\ref{mom_power_spec_mock_Verr_vs_noVerr_zevl}   indicates   the
mean values of both the monopole and quadrupole momentum power spectra
derived from massE still closely follow the intrinsic one.

\begin{figure*}
  \begin{center}
    
    \includegraphics[scale=0.85]{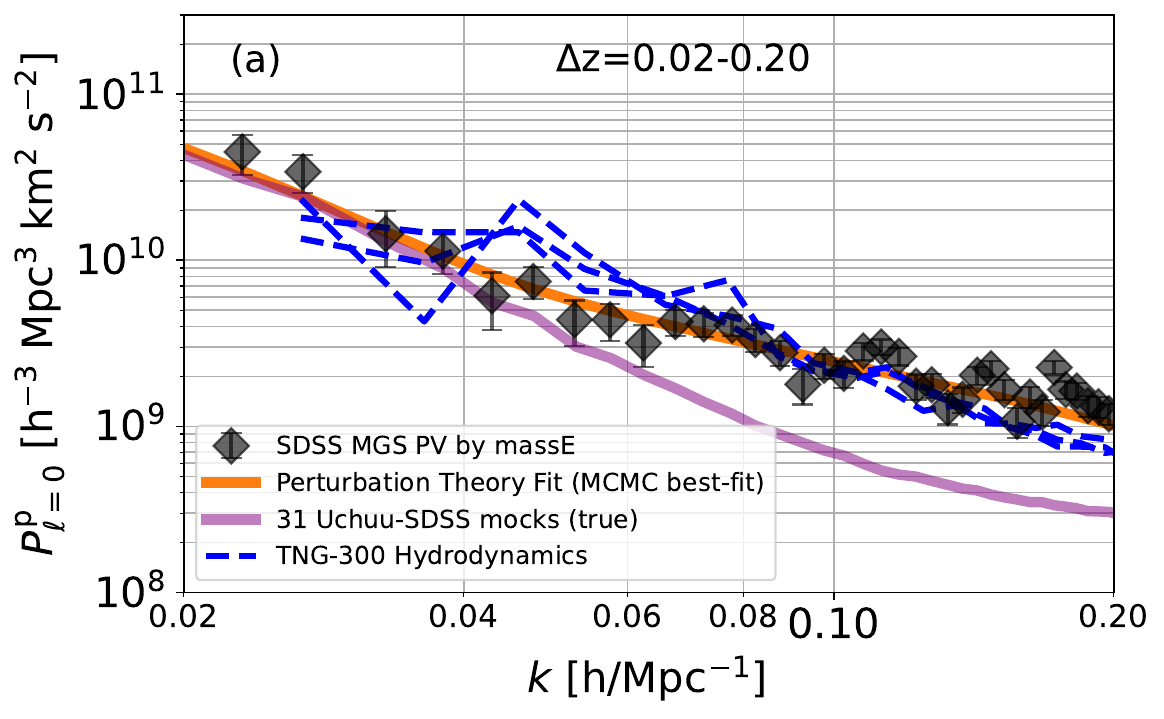}
    \hspace*{1cm}
    \includegraphics[scale=0.42]{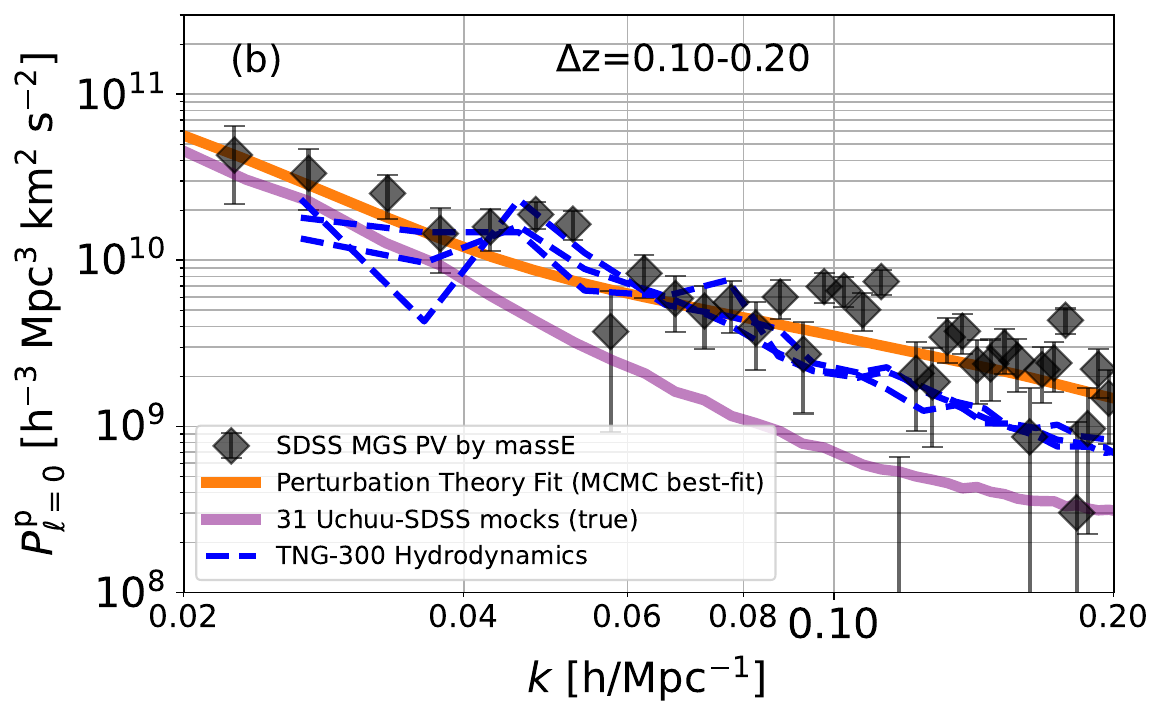}
    \includegraphics[scale=0.42]{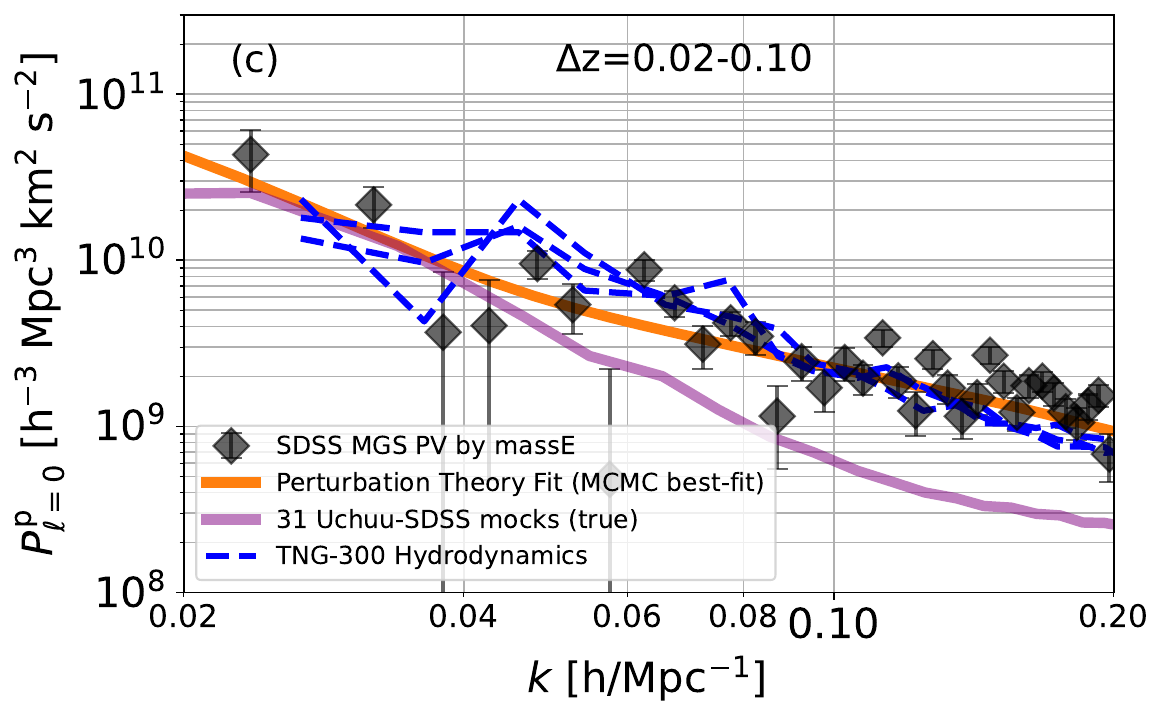}
    \caption{\label{mom_power_spec_obs_fit} {\bf (a),} the momentum power spectrum based on
      our SDSS massE-based PV catalog for the entire redshift range ${\Delta}z$=0.02-0.2. Symbols are the observed data points.
      Error bars indicate the square root of the diagonal elements of the
       covariance.
       Results of  N-body (Uchuu-SDSS) and hydrodynamic (TNG) simulations are also shown.
       {\bf (b),} the same as (a) but for the redshift range ${\Delta}z$=0.02-0.1. {\bf (c),}
       the same as (a) but for the redshift range ${\Delta}z$=0.02-0.1.} 
\end{center}
\end{figure*}

\begin{figure}
  \begin{center}
    \includegraphics[scale=0.52]{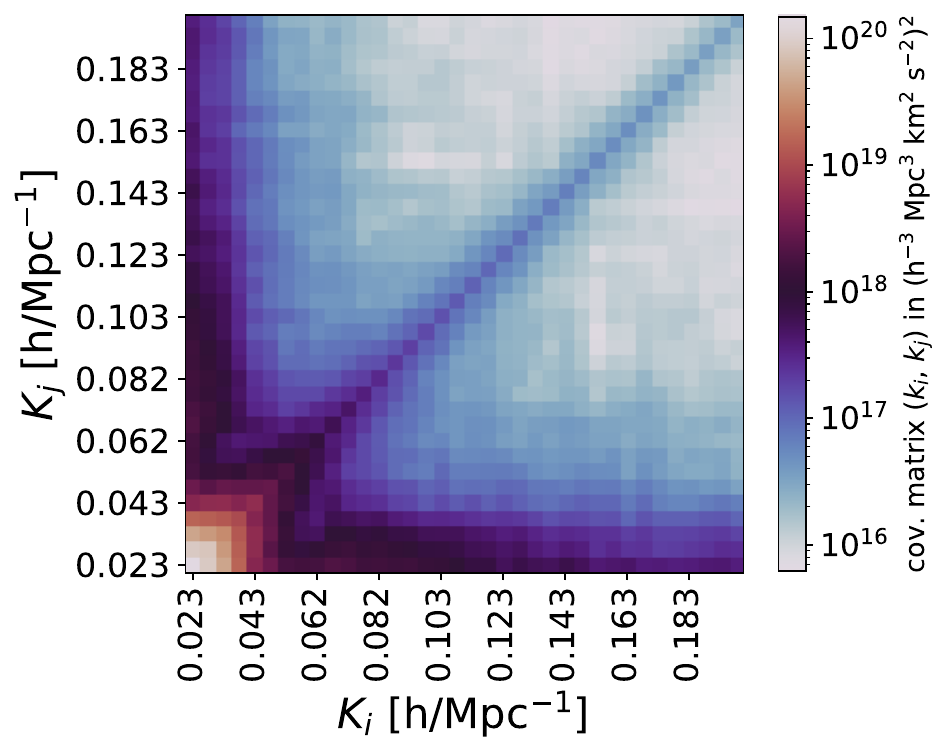}
    \caption{\label{mom_power_spec_covariance} The covariance matrix of the derived
    momentum power spectrum in Figure~\ref{mom_power_spec_obs_fit}.}
\end{center}
\end{figure}

\begin{figure}
  \begin{center}
    \includegraphics[scale=0.45]{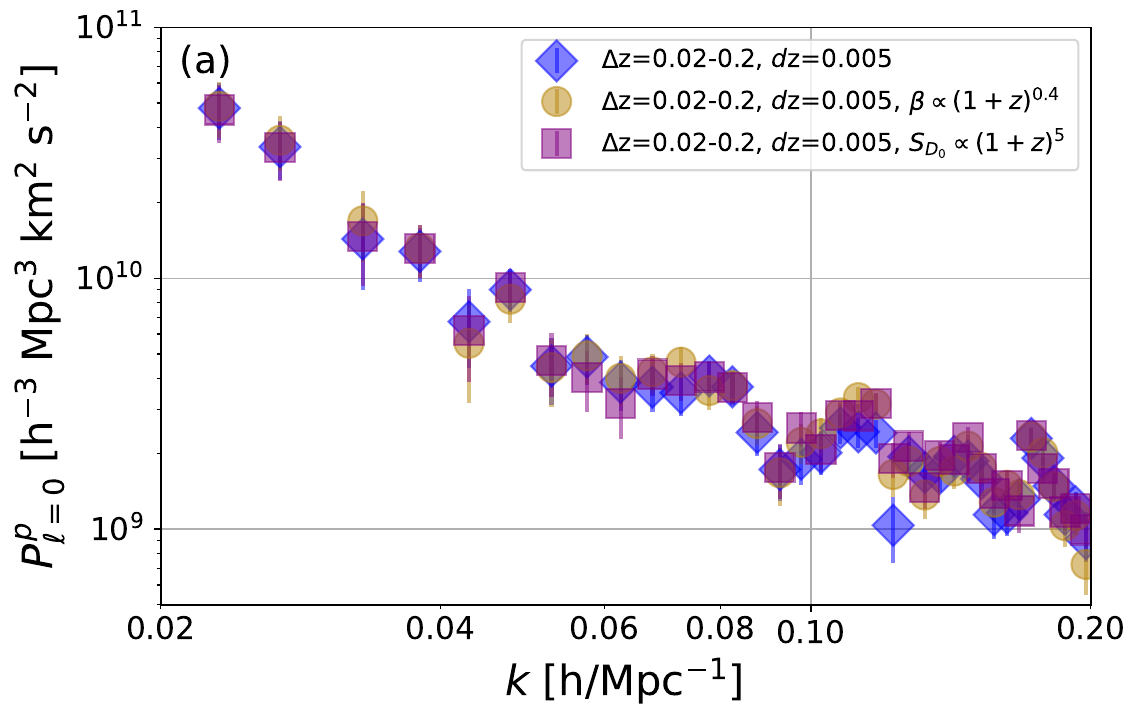}
    \includegraphics[scale=0.45]{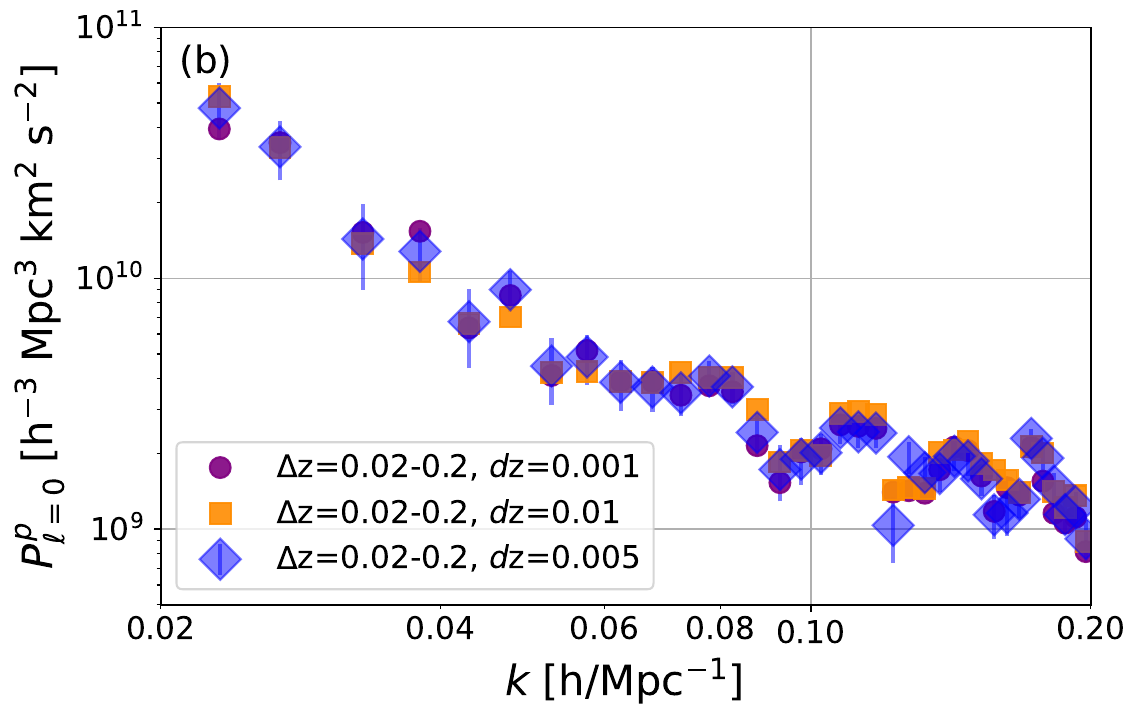}
    \caption{\label{mom_power_spec_obs_diffmassE_diffDetZ}  {\bf (a)}, the momentum power spectra
      obtained with different massE relations, including the best-fit one,
      a strong redshift evolution in the slope of the relation
      as well a strong redshift evolution in the  intercept of the relation.  {\bf (b)}, the momentum power spectrum
      obtained with three different sizes of redshift bins.}
\end{center}
\end{figure}

\begin{figure}
  \begin{center}
    \includegraphics[scale=0.60]{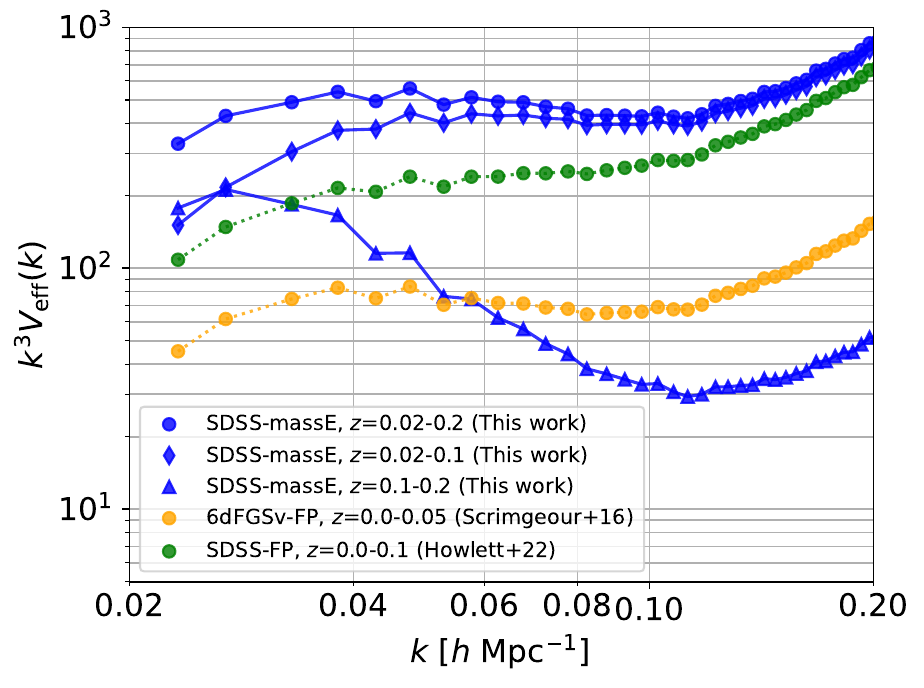}
    \caption{\label{Veff_comparison} The effective volume of our PV catalog
      as compared to the literature PV catalogs. The calculation follows
      Equation~\ref{eqn_veff}. }
\end{center}
\end{figure}

\section{The momentum power spectrum}

\subsection{The result}\label{sec_result}

Figure~\ref{mom_power_spec_obs_fit}  (a)  shows   the  result  of  the
momentum power spectrum  of our massE-based PV sample  over a redshift
range  of $\Delta$$z_{\rm  obs}$=0.02-0.2 where  PV is  measured with  a
redshift   bin   of   $dz_{\rm   obs}$=0.005.    The
covariance matrix is  shown in Figure~\ref{mom_power_spec_covariance},
which  is based  on the  result from  the mock.   As
compared  to the  true median  of  31 Uchuu-SDSS  mocks, our  observed
momentum  power   spectrum  is   slightly  higher   at  low   $k$  but
the deviation  becomes larger  at $k$  $\gtrsim$ 0.07
  $h$/Mpc.  The difference could be  the result that on small spatial
scales  non-linear  effects  affect  the  power  spectrum,  while  the
Uchuu-SDSS  simulation is  purely N-body  \citep{Ishiyama21}.  In  the
figure, we overlay the result of TNG-300 hydrodynamics simulation that
include  baryons \citep{Nelson19}.  For  this simulation,  we use  the
snapshot at $z$=0 and include all sub-halos with stellar masses larger
than  10$^{10}$ $M_{\odot}$.   The calculation  of the  momentum power
spectrum   is  obtained   through  \texttt{FFTPower}   implemented  in
\texttt{nbodykit}. Given a  specific simulation box,
  we  can have  three lines  of  sight, each  giving three  independent
  measurements, as shown  in the figure. The disparity  between TNG and
  our observation above  0.07 $h$/Mpc is now much  smaller. It remains
  unclear  what causes  the  difference between  our observations  and
  Uchuu-SDSS mock  data. The  momentum power spectrum  at high  $k$ is
  dominated  by the  product of  the density  and velocity  (e.g., see
  Fig. 13 in  \citet{Howlett19}).  The Uchuu-SDSS mock  is designed to
  have the same galaxy bias as the  observed one, a fact we confirm by
  comparing  the density  power spectra  of our  observations and  the
  Uchuu-SDSS mock.  Accordingly the difference at  high-$k$ should not
  be  attributed to  different galaxy  biases between  the two.  It is
  probable that the N-body simulation, such as Uchuu, does not account
  adequately   for non-linear  effects  as  compared with  TNG
  hydrodynamic simulation.  Fortunately, the constraint on  the growth
  rate is not particularly sensitive to the maximum wavenumber $k_{\rm
    max}$.

Figure~\ref{mom_power_spec_obs_fit}(b) and (c) presents two additional
momentum power spectra for  two sub-redshift ranges of ${\Delta}z_{\rm
  obs}$=0.10-0.20 and 0.02-0.10, respectively.  Similar to the case of
the  full  redshift  range,  the momentum  power  spectra  within  two
sub-ranges  exhibit   a  slightly  elevation  when   compared  to  the
Uchuu-SDSS result at  low $k$. However, they  show considerably larger
disparities above $k$=0.07 $h$/Mpc, but tend to match more or less the
result of the TNG hydrodynamic simulation at high $k$.


\subsection{Robustness of the measurement}

In this section, we discuss the result if varying some parameters that we use to derive
the momentum power spectrum including the massE relation itself and the redshift bin $dz_{\rm  obs}$.

In \S~\ref{sec_comp_vpec_mock}, we  confirm with mock
  data  that  the  massE-based  PV  is  insensitive  to  the  accurate
  calibration of  the massE relationship. Here  we further demonstrate
  this       with        the       observational        data.      In
Figure~\ref{mom_power_spec_obs_diffmassE_diffDetZ}(a), we  overlay two
momentum power spectra with the fiducial one: one is the case that the
normalization      $S_{D_{0}}$     of      the     massE      relation
evolves   strongly  with   redshift  as   $S_{D_{0}}$
  $\propto$  (1+$z$)$^{5}$ ,  and another  is the  one that  the slope
  $\beta$ of  the relationship evolves  strongly with the  redshift as
  $\beta$ $\propto$ (1+$z$)$^{0.4}$. For  the $\beta$, such a redshift
  evolution gives a  value at $z$=0.2 that deviates  from the observed
  one by 20-$\sigma$. In the first  case, any alteration in the value
of $S_{D_{0}}$ causes  all distances to shift uniformly  with the same
amount, thereby maintaining the offset of a distance from the Gaussian
mean.   As a  result, the  momentum  power spectrum  remains the  same
regardless of the  magnitude of the $S_{D_{0}}$ shift.   In the second
case, the momentum power spectrum changes a little bit but well within
the error  bar, which  still benefits from  our method.   For galaxies
within  a narrow  redshift  bin, their  physical properties  including
sizes, velocity dispersions and stellar masses exhibit a narrow range.
If there is a change in the  slope $\beta$, it only introduces a small
amount  of  scatter  in  the  distribution  of  distances.   Therefore
Figure~\ref{mom_power_spec_obs_diffmassE_diffDetZ}(a)     demonstrates
that  the momentum  power  spectrum  with the  massE-based  PV is  not
sensitive  to   the  accurate  calibration  of   the  massE  relation.
Figure~\ref{mom_power_spec_obs_diffmassE_diffDetZ}(b)   presents   the
momentum power spectra whose PV is measured with redshift bin sizes of
 $d{z_{\rm obs}}$=0.001 and $d{z_{\rm obs}}$=0.01, as
  compared to the fiducial one  with $d{z_{\rm obs}}$=0.005.  As shown
  in the  figure, the  three spectra are  almost the  same, indicating
  that the PV measurements is  not sensitive to $d{z_{\rm obs}}$ which
  can vary by a factor of 10.

\section{The constraint on the growth rate}

\subsection{The effective survey volume}\label{sec_eff_volume}

Before detailed modeling, we first roughly evaluate the statistical power of
our sample in constraining a cosmological parameter, given that
the signal to noise of the parameter constrained at a given $k$ is more or less
proportional to the square root of the survey effective volume at that $k$ \citep[e.g.][]{Koda14}.
Similar to the definition of the density power spectrum \citep{Feldman94, Tegmark97}, an
effective volume for the momentum power spectrum can be written as
\begin{equation}\label{eqn_veff}
  V_{\rm eff}(k) = \int \left[
    \frac{\bar{n}(z)P^{\rm P}(k)}{\langle{v_{\rm p}^{2}}\rangle_{z} + \bar{n}(z)P^{\rm P}(k)}
    \right]^{2} d^{3}r,
\end{equation}
where  $\bar{n}(z)$ and  $\langle{v_{\rm  p}^{2}}\rangle_{z}$ are  the
galaxy number density and PV variance as a function of
the  redshift,  respectively. The above equation indicates that for  high galaxy number densities
$\bar{n}(z)$ $\gg$ $\langle{v_{\rm p}^{2}}\rangle_{z}$/$P^{\rm P}(k)$,
the effective volume is just the geometrical volume and thus increases with the cube
of the redshift. On the other hand, for low galaxy
number densities, because
$\langle{v_{\rm p}^{2}}\rangle_{z}$ roughly scales with the square of the redshift at $z$ < 0.2,
the effective volume only scales linearly with the redshift.

To calculate the effective volume for  our sample, we adopt the median
of  31  Uchuu-SDSS mocks  for  $P^{\rm  P}(k)$. The  $\bar{n}(z)$  and
$\langle{v_{\rm       p}^{2}}\rangle_{z}$      are       shown      in
Figure~\ref{nofz_v2ofz}  for our  sample. We  derived
  $V_{\rm  eff}(k)$  for the  whole  redshift  range  as well  as  two
  redshift sub-ranges as  shown in Figure~\ref{Veff_comparison}, along
  with 6dFGSv  and SDSS-FP  surveys.  The 6dFGSv  data has  a spatial
coverage of 17\,000 deg$^{2}$ and a redshift range of 0.0001 to 0.0534
\citep{Scrimgeour16}.              We          use
  $(cz/(1+z){\sigma_{\eta}})^{2}$  to get  its  velocity variance  by
setting $\sigma_{\eta}$=0.324 \citep{Scrimgeour16}.  The galaxy number
density is  almost a  constant around  1.2$\times$10$^{-3}$ \unitNBAR.
We use the  same $P^{\rm P}(k)$ as for our  sample.  The SDSS FP-based
PV sample covers a redshift range of 0.0033 to 0.1 and an area of 7016
deg$^{2}$  \citep{Howlett22}.    We  use   the  above
  equation  for  the   velocity  variance  with  $\sigma_{\eta}$=0.23
\citep{Howlett22}, measure  $\bar{n}(z)$ from  their catalog  and also
adopt the median of Uchuu-SDSS mocks for $P^{\rm P}(k)$.

The         integral        quantity         $N_{\rm
    k,modes}$=$\int_{0.02}^{0.07}$$k^{2}V_{\rm eff}dk$ gives the total
  number  of $k$  modes  on linear  scales where  the  growth rate  is
  sensitive  to. As  compared to  the  6dFGSv and  SDSS FP-based  PV
catalogs,  our  sample  offers   a  factor  of  about
  \IncreaseVolumeSDSSFP\;  and  \IncreaseVolumesixdFGSv\; increase  in
  the above quantity,  respectively. As shown in the  figure, even for
  $z$<0.1,  our  effective volume  is  still  larger than  the  SDSS-FP
  catalog,  with about  a factor  of 1.7  larger in  terms of  $N_{\rm
    k,modes}$. This is because of our larger sample even at $z$ $<$ 0.1 as presented in
  \S~\ref{sec_measure_vpec}.  Note  that,
although our  sample extends to  $z_{\rm obs}$=0.2, the  galaxy number
density above $z_{\rm obs}$=0.1 of the SDSS MGS drops rapidly.

\subsection{MCMC fitting}

\begin{figure} 
  \begin{center}
    \includegraphics[scale=0.45]{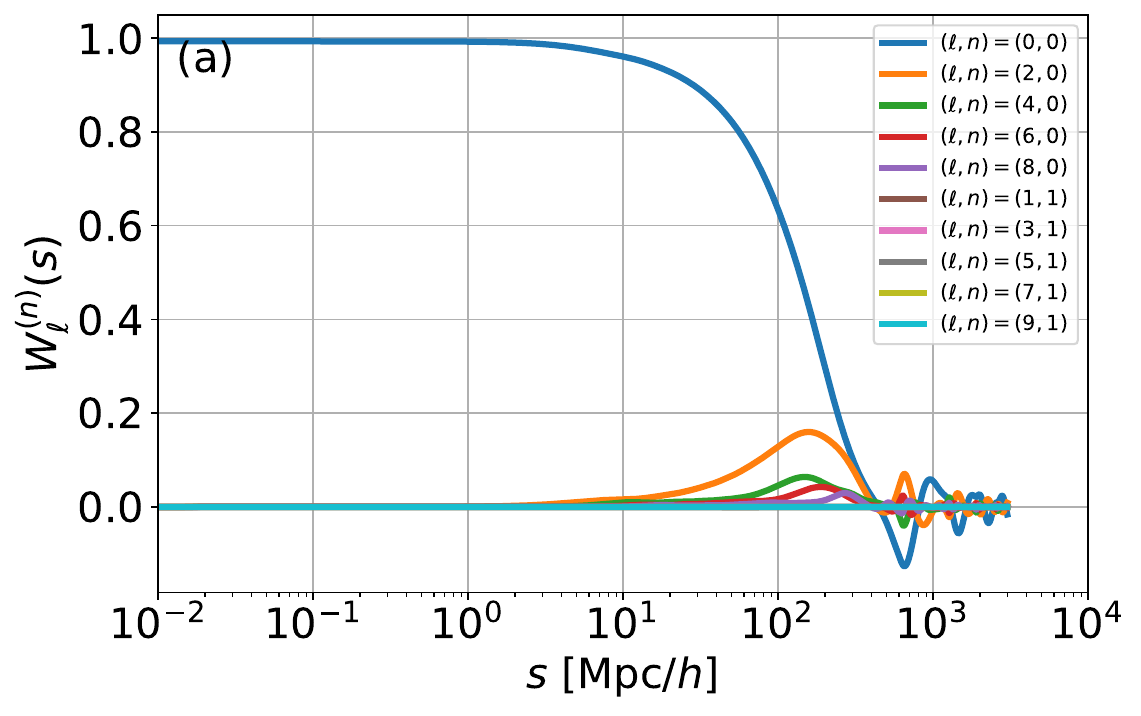}
    \includegraphics[scale=0.45]{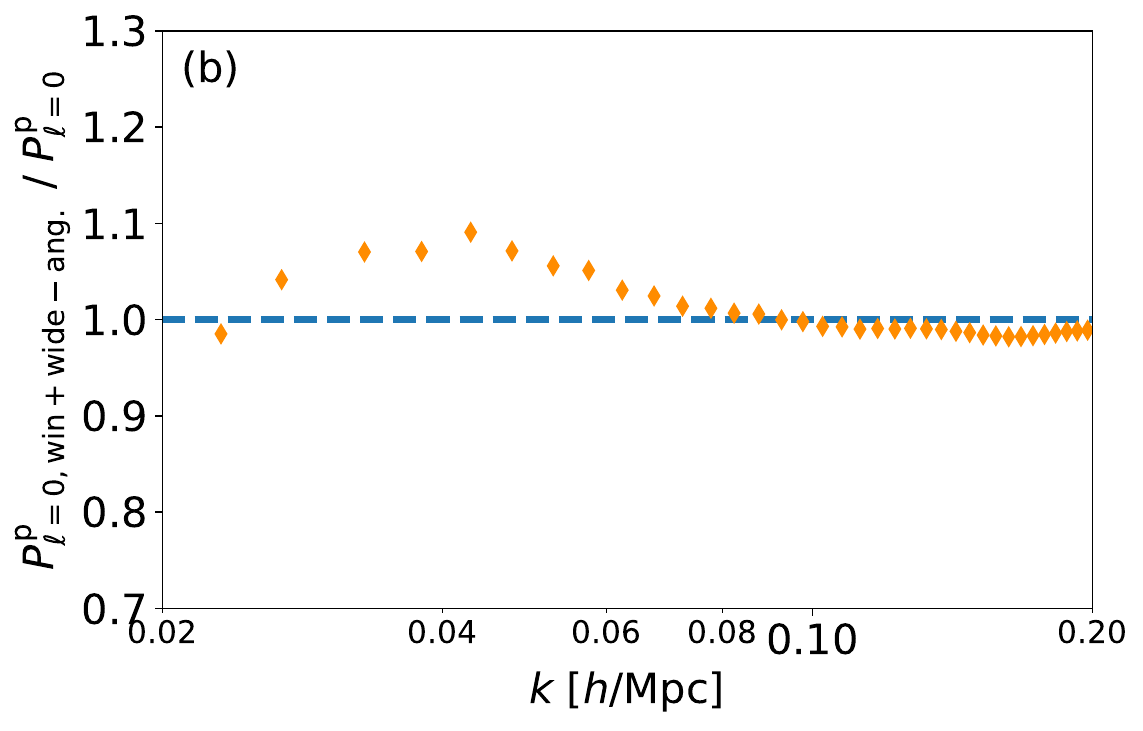}
    \caption{\label{window_function} {\bf (a,)} the window function of  our PV catalog in the
      configuration space for our sample within $\Delta$z=0.02-0.20. {\bf (b),} the ratio
      of the theoretical spectrum convolved with effects of both wide angle and window function to
      the pure theoretical spectrum. 
      The theoretical spectrum at high $k$ resolution is calculated following
      Equation~\ref{eqn_theory_ps}  at $k$ from 0.001 to 10 with a bin size of 0.005.}
\end{center}
\end{figure}

\begin{figure}
  \begin{center}
    \includegraphics[scale=0.6]{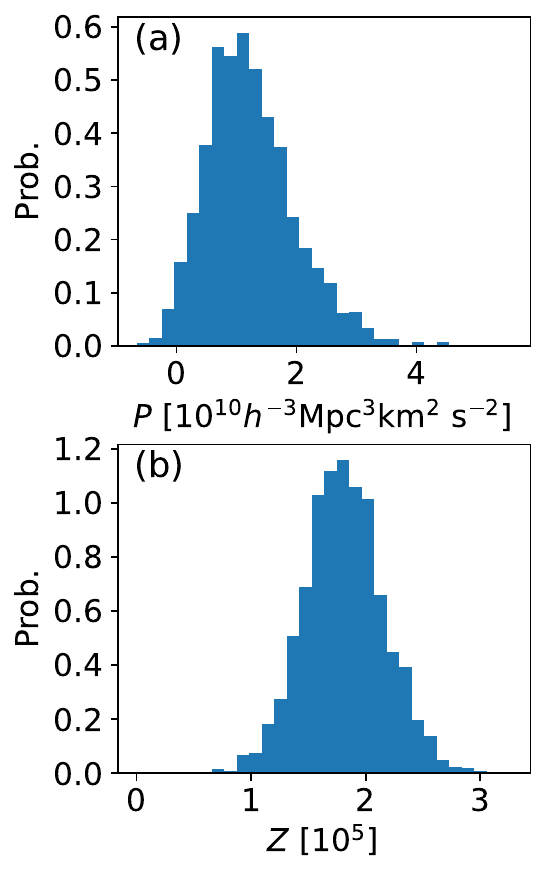}
    \caption{\label{boxcox_example}  {\bf (a),} the distribution of 1860 simulated momentum power spectra of mocks
    for $k$=[0.03, 0.04] $h$/Mpc. {\bf (b),} the Box-Cox transformed $Z$ distribution.} 
\end{center}
\end{figure}

\begin{figure}
  \begin{center}
    \includegraphics[scale=0.38]{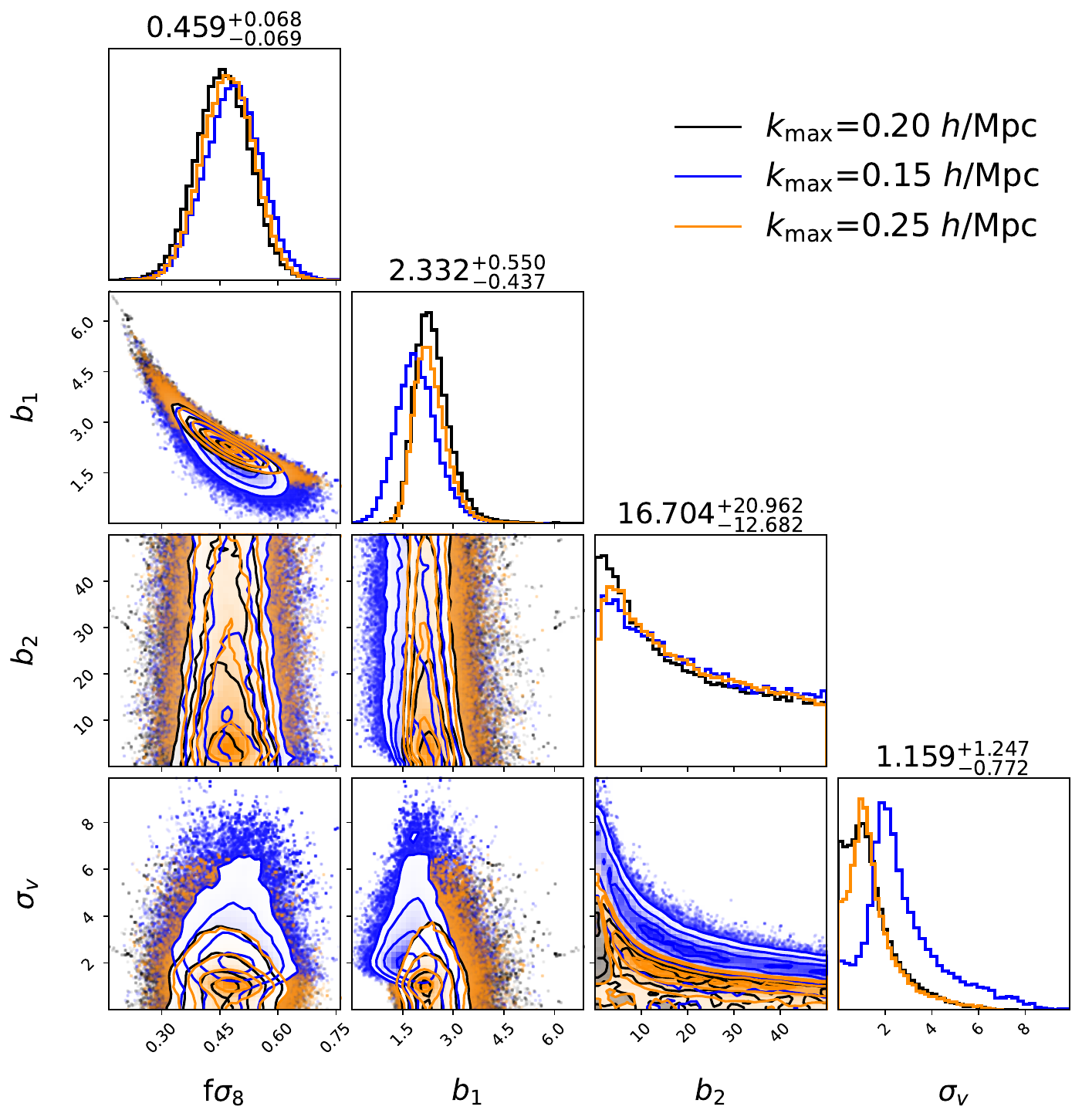}
    \caption{\label{corner_plot} The corner plot of four free parameters used to fit the observed
      momentum power spectrum. Three results corresponding to three $k_{\rm max}$ values
      are overlaid.}
\end{center}
\end{figure}

\begin{figure}
  \begin{center}
    \includegraphics[scale=0.38]{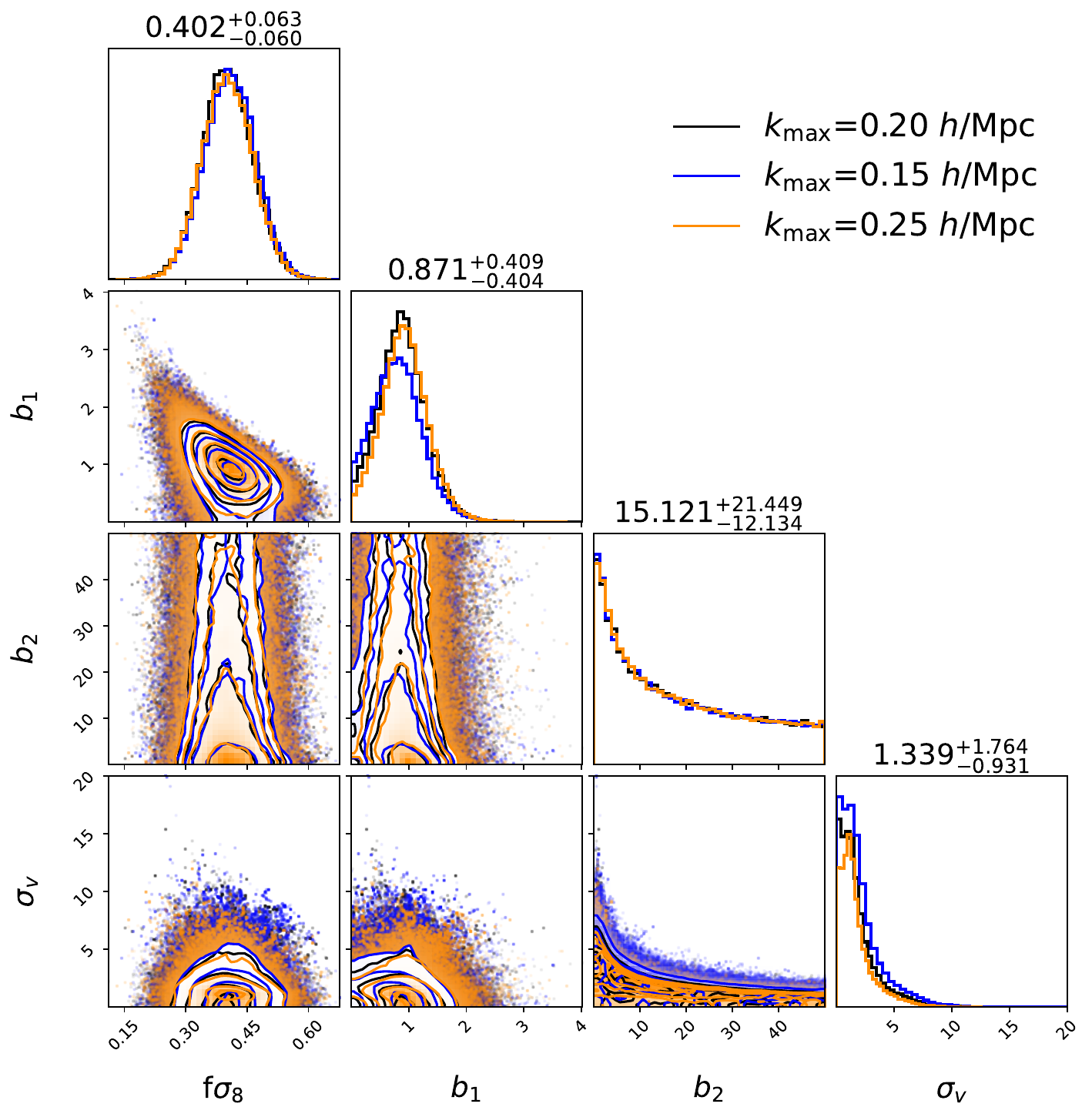}
    \caption{\label{corner_plot_mock} The same as Fig.~\ref{corner_plot} but for the fitting to mocks. }
\end{center}
\end{figure}

\begin{figure}
  \begin{center}
    \includegraphics[scale=0.45]{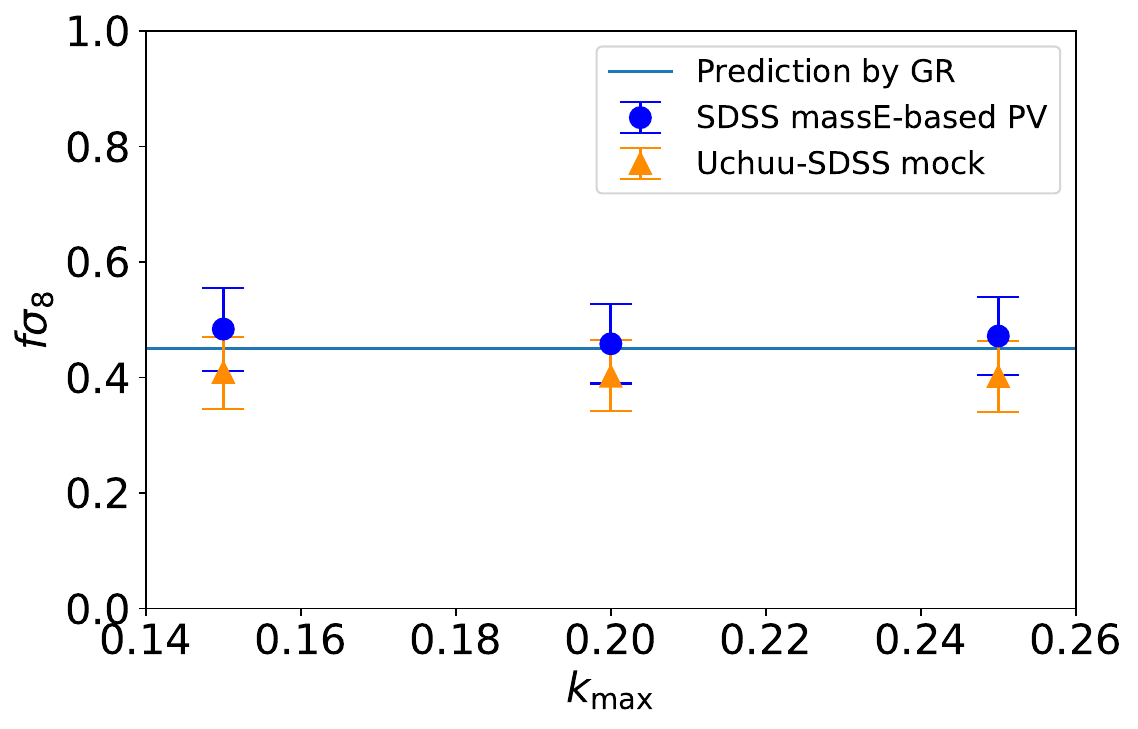}
    \caption{\label{fsigma8_diffK} The best-fit $f\sigma_{8}$ values of observations and mocks within $\Delta$z=0.02-0.2 for three
    different $k_{\rm max}$ values, where $k_{\rm min}$ is fixed at 0.02 $h$/Mpc.}
\end{center}
\end{figure}

\begin{figure}
  \begin{center}
    \includegraphics[scale=0.6]{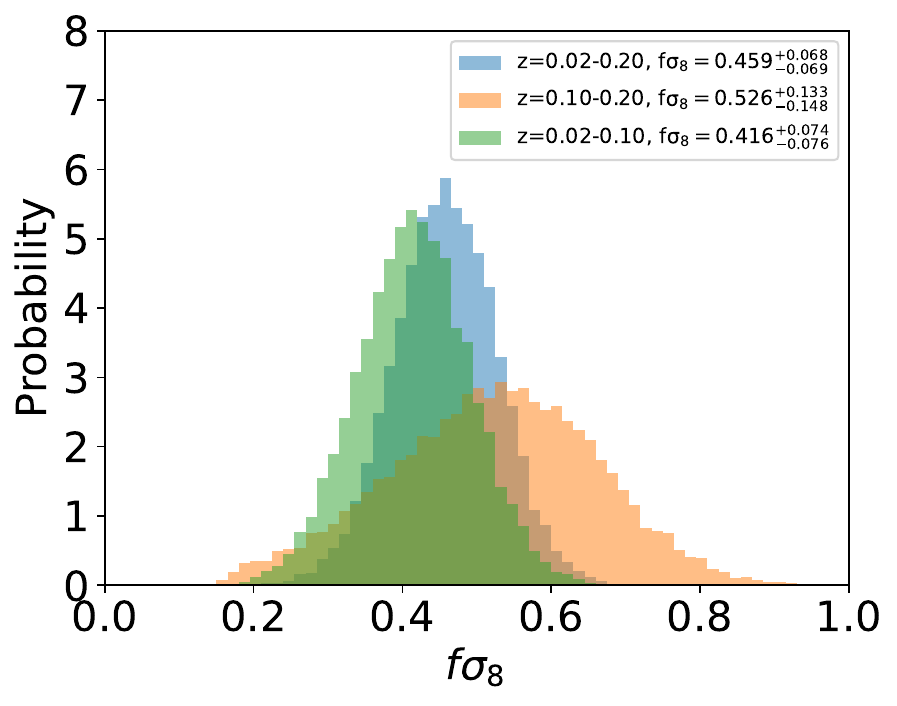}
    \caption{\label{fsigma8_diffZranges} The posterior distribution of $f\sigma_{8}$ for SDSS galaxies in
      three different redshift ranges.}
\end{center}
\end{figure}

\begin{figure*}
  \begin{center}
    \includegraphics[scale=0.6]{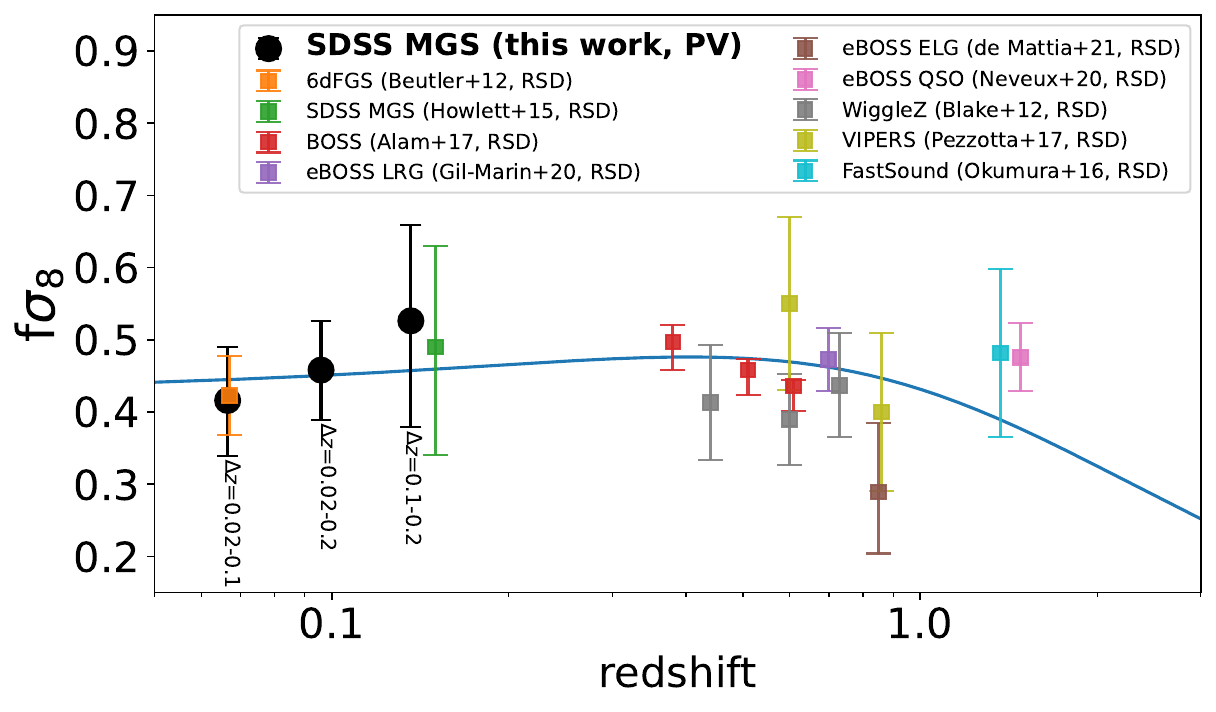}
    \caption{\label{fsigma8_asfunc_z}  Our measurement of the growth rate with the momentum power spectrum overlaid
      on those obtained with redshift space distortion including 6dFGS \citep{Beutler12}, SDSS MGS
      \citep{Howlett15}, BOSS \citep{Alam17}, eBOSS LRG \citep{Gil-Marin20},
      eBOSS ELG \citep{deMattia21}, eBOSS QSO \citep{Neveux20}, WiggleZ \citep{Blake12},
      VIPERS \citep{Pezzotta17} and FastSound \citep{Okumura16}. The solid line is the prediction
    by the GR under the Planck 2018 cosmology.}
\end{center}
\end{figure*}

We  use the  model  that is  based on  the  distribution function  and
Eulerian perturbation  theory as  developed by  \citet{Vlah12, Vlah13,
  Okumura14} and \citet{Saito14}. The analytic formula of the model is
compiled  in  the  appendix  of \citet{Howlett19}. Here we give the basic one:
\begin{equation}\label{eqn_theory_ps}
P^{P}_{\ell}(k) = (2\ell+1) \int^{1}_{0}P^{P}(k,\mu)L_{\ell}(\mu)d\mu,
\end{equation}
where
\begin{equation}
 P^{P}(k,\mu) = (aH)^{2}k^{-2}(P_{11}+\mu^2(2P_{12}+3P_{13}+P_{22})).
\end{equation}
The term $P_{11}$, $P_{12}$, $P_{13}$ and $P_{22}$ can be calculated from the linear power spectrum while
containing free parameters such as the growth rate $f$, the
linear-scale bias $b_{\rm 1}$, the non-linear bias $b_{\rm 2}$, the
non-linear  velocity dispersion  $\sigma_{v}$ etc. Through these parameters the model
accounts for the effect  of  small-scale non-linear  motion  on the  power
spectrum.

Before running the  MCMC fitting, we first quantify the  effect of the
window function  for the momentum  power spectrum measurement.  Due to
the survey  volume as well  as the sample  selection as a  function of
both spatial positions  and redshift, the power  spectrum suffers from
the effect of  the window function. We evaluate this  effect using the
python                                                            code
\texttt{pypower.CatalogSmoothWindow}\footnote{https://github.com/cosmodesi/pypower}
with following steps: (1) a random elliptical catalog: we start with a
random      catalog      available      for     the      SDSS      LSS
sample\footnote{http://sdss.physics.nyu.edu/vagc/lss.html,
http://sdss.physics.nyu.edu/lss/dr72/}  \citep{Blanton05}.   We  first
follow  the observed  fraction of  ellipticals  in the  LSS sample  to
randomly  pick objects  from the  above catalog  to define  the random
elliptical catalog.  By fitting  the observed redshift distribution of
the MGS-elliptical sample with multiple Gaussian profiles, we assign a
random redshift to those in the random elliptical catalog. After that,
we remain those  in the North Cap region and  limit the redshift range
to  0.02-0.2.  (2)  the window  function:  based on  the above  random
catalog,  we then  use \texttt{pypower.CatalogSmoothWindow}  to derive
multipoles of the  window function up to $\ell$=9 with  the wide angle
up to  the order of  $n$=1. The result  in the configuration  space is
shown  in  Figure~\ref{window_function}(a) for ${\Delta}z$=0.02-0.20.    (3)  the  theory  power
spectrum:   we  then   use  Equation~\ref{eqn_theory_ps}   to  produce
theoretical momentum power spectrum at $\ell$=0,  2 \& 4 with $k$ from
0.001 to  10 with a  resolution of 0.005.  (4) the power  spectrum with
window function and wide angle: we then convolve the above theoretical momentum power
  spectrum  with the window  function and wide-angle, and show its ratio to the pure
  theoretical spectrum in Figure~\ref{window_function}(b). It is found
that   the  effect   of  the   window  function   and  wide-angle   is
small but a difference of a few to ten percent is seen 
 for $k$ values between
  0.02 and  0.1 $h$/Mpc.  We thus  carry out  the MCMC  fitting
  by incorporating correction factors into the theoretical power spectrum. The correction factors for two redshift
  sub-ranges of ${\Delta}z$=0.02-0.10 and ${\Delta}z$=0.10-0.20 are also produced following the same procedure,
  and their differences from the above one are small.

As discussed in \citet{Qin19}, the distribution of the momentum power
  spectrum at given $k$ is not
always normal as shown in Fig.~\ref{boxcox_example}(a), and it is thus suggested
to first Box-Cox transform \citep{Box64} the momentum power spectrum to $Z$ for which the MCMC fitting
is carried out:
\begin{equation}
 Z = [(P+P_{\rm shift})^{\lambda}-1]/\lambda
\end{equation}
Here, because of large errors in $P(k)$ some P($k$) is negative
so that we add a $P_{\rm shift}$ which is set to be the minimum negative
value of the distribution. The code \texttt{scipy.stats.boxcox} is used to find the best $\lambda$
for each transformation.

We set the background cosmological parameter to the Planck 2018 result
with  $\sigma_{8}$($z$=0)=0.811 \citep{Planck20},  based on  which the
linear  power spectrum  at  $z$=0 is  calculated as  an  input of  the
theoretical power  spectrum through \texttt{nbodykit} using  the CLASS
transfer function.  In this case, if  we multiply the derived $f$ with
$\sigma_{8}$($z$=0),  it  gives   $f\sigma_{8}$  at  $z=z_{\rm  eff}$.
Following \citet{Qin19},  we set  the following four  free parameters:
$f$, $b_{\rm 1}$, $b_{\rm 2}$  and $\sigma_{v}$.  The MCMC fitting has
been    carried    out     through    Python    code    \texttt{emcee}
\citep{Foreman-Mackey13}.   Priors of  four parameters  are listed  in
Table~\ref{tab_fit_result}.

We  set $k_{\rm  min}=0.02$ $h$/Mpc  and conduct  the
  MCMC  fitting  for $k_{\rm  max}$  values  of  0.15, 0.20  and  0.25
  $h$/Mpc, respectively, for  galaxies within $\Delta$z=0.02-0.2.  The
  corner   plots    of   three   fitting   results    are   shown   in
  Figure~\ref{corner_plot}.  The growth  rate $f\sigma_{8}$  remains
unaffected  by the  non-linear  parameters  $b_{2}$ and  $\sigma_{v}$.
 While   there   are  some   correlations   between
  $f\sigma_{8}$ and  $b_{1}$ at  a given  $k_{\rm max}$,  the best-fit
  values of $f\sigma_{8}$ exhibit no  sensitivity to $k_{\rm max}$.  
This is consistent  with the expected advantage of  the momentum power
spectrum  that  is   affected  little  by  the   bias  and  non-linear
effects. Figure~\ref{fsigma8_diffK}  offers a further
  comparison  in  $f\sigma_{8}$  between observations  and  Uchuu-SDSS
  mocks, revealing that our observations  and the mock data conform to
  the predictions  of General  Relativity (GR)  and display  no trends
  with  respect to  $k_{\rm  max}$. The  errors  from observation  are
  slightly larger  than those from  mocks. This is  reasonable because
  the  median momentum  power spectrum  of  mocks at  each $k$  almost
  precisely  lies   on  the  theoretical  prediction.    As  shown  in
  Figure~\ref{fsigma8_diffZranges}, the best fit is slightly below the
  median of data points, which is caused by the fact that the covariance matrix
  contains  significant non-diagonal  parts. This  is also  the reason
  that  the best-fitted  $f\sigma_{8}$ of  mocks are  not exactly  but
  slightly smaller than the GR's prediction.  

    As        shown       in        Figure~\ref{corner_plot}       and
    Figure~\ref{corner_plot_mock}, there  are noticeable discrepancies
    in  the best-fit  $b_{1}$  between observations  and mocks.   This
    discrepancy is  indicative of  a disparity  in the  momentum power
    spectrum  at  high  $k$  regime   between  the  two  as  shown  in
    Figure~\ref{mom_power_spec_obs_fit}.  While  both observations and
    mocks have  a similar galaxy  bias ($\sim$  1.5) as seen  by their
    density  power spectrum,  the difference  in the  above result  is
    likely due to the velocity field on small scales. As stated above,
    the  comparison   between  Uchuu-SDSS   and  TNG   indicates  that
    Uchuu-SDSS  as  a N-body  simulation  may  not fully  account  for
    non-linear effects. In addition, the observations  exhibit higher values than
    TNG above  $k$= 0.1 $h$/Mpc  by about  20-30\%.  There may  be two
    reasons  for this:  one is  that there  is some  random zero-point
    offsets in velocity similar to those seen in mock simulations (see
    Figure~\ref{vmock_vtrue}), which adds a constant to the shot noise
    \citep{Howlett19}.   This   addition  predominantly   affects  the
    high-$k$ regime but becomes negligible at low-$k$; another is that
    since  the   mocks  underestimate  the  power   at  high-$k$,  the
    corresponding      error       at      high-$k$       is      also
    underestimated. Nevertheless, these factors  should not affect our
    conclusion regarding $f\sigma_{8}$, as  it primarily relies on the
    low $k$ regime.
  
We  also  carry  out  fit  to   the  momentum  power  spectra  of  two
sub-redshift  ranges of  ${\Delta}z_{\rm  obs}$=0.02-0.1 and  0.1-0.2,
which gives $f\sigma_{8}$ of  \fsigmaLZ\, and \fsigmaHZ, respectively,
for   $k_{\rm  max}$=0.20   $h$/Mpc.   As  shown   in
  Figure~\ref{fsigma8_diffZranges},  they  are   consistent  with  the
  result of  the entire redshift  range, which yields  a $f\sigma_{8}$
  value of  \fsigma, well  within the 1-$\sigma$  confidence interval.
  The inverse variance  of $f\sigma_{8}$ for the  whole redshift range
  is almost  the same as the  sum of those of  two redshift sub-ranges
  too.  These results offer additional  evidence for the robustness of
  our results.

     Table~\ref{tab_fit_result}           lists
  $\chi^{2}$/d.o.f.  values  for  three  redshift  ranges,  which  are
  between 2 and  3. We suspect that  this is due to the  fact that the
  mock data  under-estimate the  momentum power spectra  and associated
  errors at high $k$ values as stated in \S~\ref{sec_result}. To test this for $\Delta{z}$=0.02-0.2, we
  increase the error by a factor of  1.5 for $k$ above 0.1 $h$/Mpc and
  corresponding  non-diagonal parts  in  the covariance  matrix. 
    By carrying out the MCMC fitting again, it  is
  found  that the  best-fit  $f\sigma_{8}$ remains  unchanged but  now
  $\chi^{2}$/d.o.f.   is 36/32.  This  further  demonstrates that  the
  growth rate measured from momentum  power spectrum is insensitive to
  momentum power spectra on small scales.  

\subsection{Comparisons with other studies}

\citet{Qin19} adopted the same theoretical model with
  the  same  set of  free  parameters for  the  6dFGSv and  2MTF  PV
data. They  obtained $f\sigma_{8}$=0.404$^{+0.082}_{-0.081}$  with the
combined density  power spectrum  and momentum  power spectrum  of two
data-sets together.This constraint is similar to that
  by \citet{Turner23} using the combined  velocity and density data of
  the 6dFGSv survey.  If adopting the momentum power  spectrum of the
6dFGSv survey alone, \citet{Qin19}  estimated 68\% confidence range to
be 0.226  (see their Table 2).   Our measurement from
  ${\Delta}z$=$0.02-0.20$  offers a  factor  of  1.7 higher  accuracy.
  This improvement in accuracy is  somewhat less than that expected  by the square
  root    of     the    increase     in    the     effective    volume
  ($\times$\IncreaseVolumesixdFGSv),   but   are   consistent   within
  30\%. This is reasonable given different mocks or fitting techniques \citep{Eisenstein05}.
  \citet{Lai23} measured  the growth  rate using  the maximum-likelihood
  fields method through both density and PV data of the SDSS-FP survey
  \citep{Howlett22}.      The     derived      growth     rate      is
  0.405$^{+0.076}_{-0.071}$(stat)$\pm$0.009  (sys),  whose  confidence
  range   is  consistent   with   our  low-$z$   one  (\fsigmaLZ\,   for
  ${\Delta}z$=0.02-0.1). They  did not  report the  result when  only using PV
  data. However, as mentioned  earlier, although both our catalog and
  the SDSS-FP catalog are based on  the same parent sample, our massE-based
  PV   catalog  includes  \IncreaseFPnum\,  more   objects  than   the
  SDSS-FP catalog at $z$ < 0.1.  This  could account for   our  PV-only result exhibiting a
  similar level of error as theirs of combined PV and density data.
  \citet{Saulder23} have provided the predicted number density and PV error
  for the DESI survey, from which they estimated a 68\% confidence range for
  $f\sigma_{8}$, which is 0.126 and 0.092 for $k_{\rm max}$ values of
  0.1 $h$/Mpc and 0.2 $h$/Mpc, respectively. We measure their effective
  volume and find that their number of $k$-modes
  between 0.02 and 0.07 $h$/Mpc is 1.7 times more than our entire redshift
  range. This corresponds to a 1.3-fold enhancement in their precision,
  consistent with our constraint of $f\sigma_{8}$ at a value of \fsigma\, at $k_{\rm max}$=0.2 $h$/Mpc.

\citet{Howlett15} measured the redshift space distortion of the SDSS MGS sample through
two-point correlation function, resulting in a derived $f\sigma_{8}$ of 0.49$^{+0.15}_{-0.14}$. Our
massE-based PV measurement of the same sample thus offers
a factor of \IncreaseErrorRSD\, improvement in the error of $f\sigma_{8}$. In Figure~\ref{fsigma8_asfunc_z},
we overlay our measurement on most recent collection of $f\sigma_{8}$ that has been measured with
redshift space distortion \citep{Alam21}.

\begin{table*}
\begin{center}
\caption{\label{tab_fit_result} The parameters of the MCMC fitting to the momentum power spectrum in three redshift ranges.}
\begin{tabular}{llllllllllllll}
\hline
par. name    & priors &  ${\Delta}z$=0.02-0.1 & ${\Delta}z$=0.1-0.2  & ${\Delta}z$=0.02-0.2\\

\hline

($k_{\mathrm{min}}$, $k_{\mathrm{max}}$)    &    & (0.02, 0.20)  & (0.02, 0.20)   & (0.02, 0.20)  \\
$z_{\mathrm{eff}}$ &        & 0.0666                 &         0.1363        &     0.0959      \\
$f\sigma_{8}$ (GR) &        & 0.4448                 &         0.4574        &     0.4505      \\
\hline
$f\sigma_{8}$ & U(0, 5)     & 0.416$^{+0.074}_{-0.076}$ & 0.526$^{+0.133}_{-0.148}$ & 0.459$^{+0.068}_{-0.069}$\\ [4pt]
$b_{1}$       & U(0, 10)      & 2.36$^{+0.68}_{-0.50}$ & 2.68$^{+1.40}_{-0.82}$ & 2.33$^{+0.55}_{-0.44}$\\ [4pt]
$b_{2}$       & U(0,50)     & 16.32$^{+20.73}_{-12.52}$ & 15.17$^{+21.47}_{-11.92}$ & 16.68$^{+21.04}_{-12.63}$\\ [4pt]
$\sigma_{v}$[$h$/Mpc] & U(0,inf)    & 1.28$^{+1.42}_{-0.83}$ & 1.20$^{+1.61}_{-0.84}$ & 1.16$^{+1.25}_{-0.79}$\\ [4pt]
$\chi^{2}$/d.o.f. &   &   75.5/31  &  91.1/32  &  65.5/32\\

\hline
\end{tabular}\\
U stands for a uniform distribution. The error of the best fit is given for
the 68\% confidence range.
\end{center}

\end{table*}


\begin{figure}
  \begin{center}
    \includegraphics[scale=0.45]{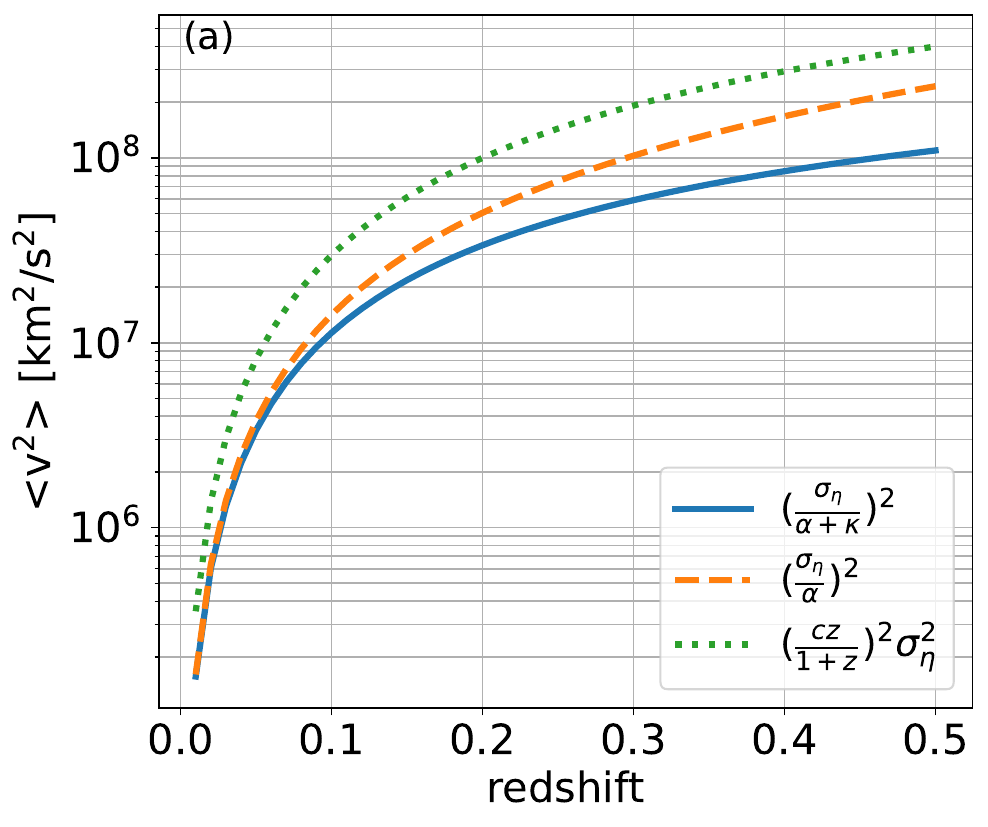}
    \includegraphics[scale=0.45]{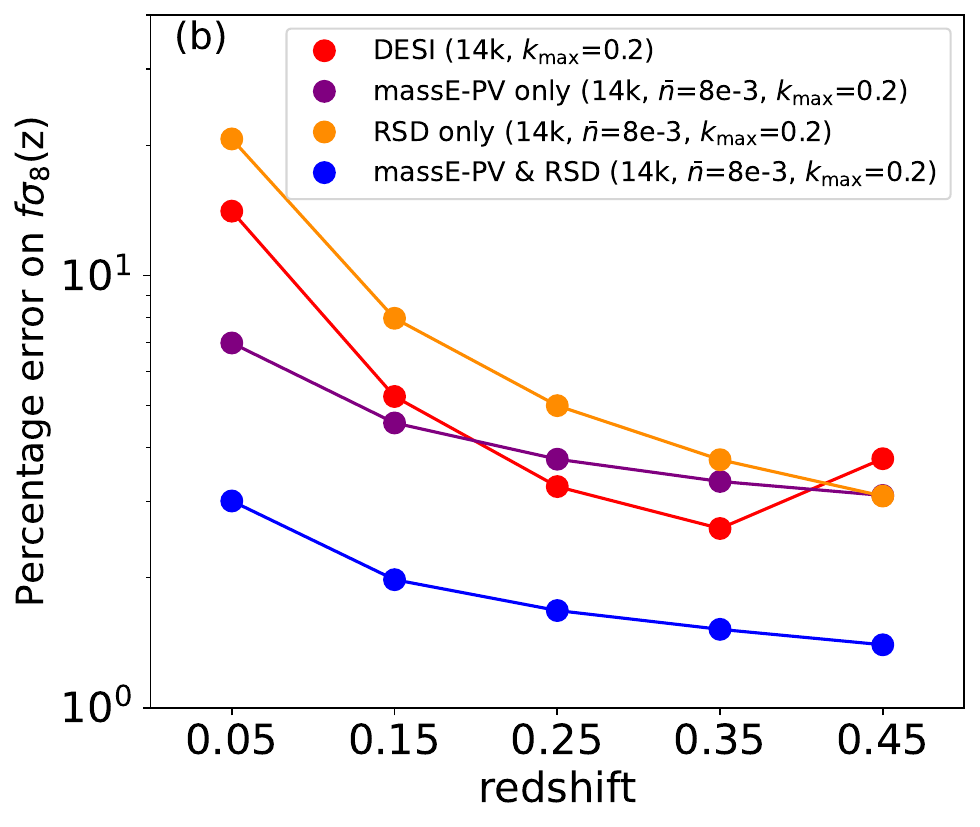}
    
    \caption{\label{forecast}  {\bf (a),}  the error
        variance  of  PV  as  a function  of  redshift  for  different
        formulae.        The       massE-based        PV       follows
        $\sigma_{\eta}^{2}/(\alpha+\kappa)^{2}$, where $\sigma_{\eta}$ is
        the distance error of ln($D_{\rm c}$), $\alpha$ and $\kappa$ are defined
        in Equation~\ref{eqn_vp_eta}.    {\bf   (b),}   the
        Fisher matrix forecast for the constraint on $f\sigma_{8}$ for
        a survey with an area of 14 k square degree and average galaxy
        number density of 8$\times$10$^{-3}$  $h^{3}$/Mpc$^{3}$.  The DESI's forecast
        is from \citet{DESI16}. }
    
\end{center}
\end{figure}

\subsection{Fisher-matrix forecast for the massE-based PV}

Our work  has demonstrated that  the massE can  serve as a  new cosmic
ruler with only  two nuisance parameters, allowing for  the probing of
the  PV  beyond  the  current  redshift  limit  of  0.1.   For  future
perspective, we  envision that  there is  no apparent  limit for  the PV
measurement with the massE to higher redshift, as it is insensitive to
the   accurate    calibration   of   the   massE    relation   itself.
In   addition,   because    of   $\kappa$   term   in
  Equation~\ref{eqn_vp_eta}  that  is  used to  convert  the  distance
  measurement to PV,  the PV variance scales with  the redshift slower
  than $z/(1+z)$ by a few times, especially above $z$=0.2, as shown in
  Figure~\ref{forecast}(a).   As  a  result,  although  measuring  the
  momentum power spectrum still requires  a high galaxy number density
  above $z$=0.2, it  has a potential to provide  strong constraints on
  the growth rate in the late universe where dark energy dominates. To
  illustrate this  quantitatively, we  use the  Fisher-matrix forecast
  code\footnote{https://github.com/CullanHowlett/PV\_fisher}
  \citep{Howlett17a,  daCunha17} with  a survey  area of  14 k  square
  degree,    a   distance    error   of    $\sigma_{\eta}$=20\%, a galaxy number density of
  8$\times$10$^{-3}$ $h^{3}$/Mpc$^{3}$ and  Equation~\ref{eqn_vp_eta}   for   the   PV  error.   As   shown   in
  Figure~\ref{forecast}(a), PV-only  data yields  stronger constraints
  compared to  the RSD only for  this specific data-set where $k_{\rm max}$ is set to be 0.2 $h$/Mpc. On  the other
  hand, the momentum power spectrum  mainly relies on linear scales to
  extract  cosmological  information,   complementing  the  RSD.   The
  combination  of  these two  offers  significant  improvement in  the
  growth  rate,  potentially exceeding  the  DESI's  constraints by  a
  factor of 1.7 to 4.7 in terms of errors.

\section{Conclusions}

In this study we utilize the massE relation of galaxies as a new cosmic ruler to
estimate the PV of  SDSS MGS elliptical galaxies from $z$=0.02
out to $z$=0.2, 
and measure their momentum power spectrum. The main conclusions are as followings: 

(1) PV is measured for a galaxy based on the offset of its massE-based distance
from the Gaussian mean of the distance distribution in a narrow redshift bin.

(2) The final
PV catalog consists of  \totnumPVCatalog\; objects, which is almost seven times
larger than the previously
largest catalog. The redshift coverage is well beyond the limit of around 0.1 in literature
studies using other distance rulers such as Tully-Fisher and FP. 

(3) By mimicking the PV measurements with our method in Uchuu-SDSS mock galaxies, we show
that both the PV of individual galaxies and the momentum power spectra can be recovered well, and both are insensitive to accurate calibration of the massE relation itself.

(4) We derive the momentum power spectrum based on the massE-based PV catalog over spatial scales
from 0.02 to 0.2 h/Mpc. The spectrum shows little dependence on
the massE relation, including its slope and intercept. The derived spectra for two sub-redshift ranges of
0.02-0.10 and 0.10-0.20 are consistent with the one over the whole redshift range.
The spectrum is also insensitive to the size of the redshift bin in which PV is measured. These demonstrate
the robust of the massE-based PV measurement.

(5) By fitting perturbation theory model to the derived momentum power spectrum with four free parameters,
we find the best-fit $f\sigma_{8}$ to be \fsigma\, at $z_{\rm eff}$=\zeff. Its error is \IncreaseErrorRSD\, times smaller than the value
based on the redshift space distortion of the SDSS MGS.

(6)  The massE-based  PV is  insensitive to  accurate
  calibration of the massE relation  itself. This combined with a slow
  increase in  the PV  errors with redshift  beyond $z$=0.2,  makes the
  massE-based  PV   a  potentially  competitive  new   method  to  offer
  high-precision constraints  on the growth  rate, in addition  to the
  RSD. 

\section*{Acknowledgements}

We  thank  the anonymous referee  for  detailed and constructive  comments  that
improved the manuscript significantly. Y.S. acknowledges  the support from  the National Key R\&D  Program of
China  No. 2022YFF0503401,  the National  Natural
Science  Foundation   of  China   (NSFC  grants   11825302,  12141301,
12121003, 12333002),  and the  New  Cornerstone Science  Foundation through  the
XPLORER PRIZE.

Funding for  the SDSS and SDSS-II  has been provided by  the Alfred P.
Sloan Foundation, the Participating Institutions, the National Science
Foundation, the  U.S.  Department of Energy,  the National Aeronautics
and Space Administration, the  Japanese Monbukagakusho, the Max Planck
Society, and  the Higher  Education Funding  Council for  England. The
SDSS  Web Site  is http://www.sdss.org/.  The SDSS  is managed  by the
Astrophysical    Research    Consortium    for    the    Participating
Institutions. The  Participating Institutions are the  American Museum
of  Natural History,  Astrophysical Institute  Potsdam, University  of
Basel,  University  of  Cambridge, Case  Western  Reserve  University,
University of Chicago, Drexel  University, Fermilab, the Institute for
Advanced  Study,   the  Japan   Participation  Group,   Johns  Hopkins
University, the  Joint Institute  for Nuclear Astrophysics,  the Kavli
Institute  for   Particle  Astrophysics  and  Cosmology,   the  Korean
Scientist Group, the Chinese Academy  of Sciences (LAMOST), Los Alamos
National  Laboratory, the  Max-Planck-Institute for  Astronomy (MPIA),
the  Max-Planck-Institute for  Astrophysics  (MPA),  New Mexico  State
University,   Ohio  State   University,   University  of   Pittsburgh,
University  of Portsmouth,  Princeton  University,  the United  States
Naval Observatory, and the University of Washington.

\section*{Data Availability}

The data produced here are available upon reasonable request.
 



\bibliographystyle{mnras}
\bibliography{ms} 




\appendix




\bsp	
\label{lastpage}
\end{document}